\documentclass[aos,preprint]{imsart}
\usepackage{subfigure}
\usepackage[figuresright]{rotating}
\usepackage{bm}
\RequirePackage{amsthm,amsmath,amsfonts,amssymb}
\RequirePackage[authoryear]{natbib} 

\startlocaldefs
\numberwithin{equation}{section}
\theoremstyle{plain}
\newtheorem{thm}{Theorem}[section]
\newtheorem{remark}{Remark}[section]

\newtheorem{pro}{Proposition}[section]

\endlocaldefs

\begin{document}

\begin{frontmatter}
\title{Root $n$ consistent extremile regression and its supervised and semi-supervised learning}
\runtitle{Root $n$ consistent extremile regression}

\begin{aug}
\author[A]{\fnms{Rong} \snm{Jiang}\ead[label=e1]{jrtrying@126.com}},
\and
\author[B]{\fnms{Keming} \snm{Yu}\ead[label=e2,mark]{keming.yu@brunel.ac.uk}}
\address[A]{Shanghai Polytechnic University,  \printead{e1}}

\address[B]{Brunel University London,  \printead{e2}}
\end{aug}

\begin{abstract}
Extremile \citep{r1} is a novel and coherent measure of risk, determined by weighted expectations rather than tail probabilities. It finds application in risk management, and, in contrast to quantiles, it fulfills the axioms of consistency, taking into account the severity of tail losses. However, existing studies \citep{r1,r41}  on extremile involve unknown distribution functions, making it challenging to obtain a $\sqrt{n}$-consistent estimator for unknown parameters in linear extremile regression. This article introduces a new definition of linear extremile regression and its estimation method, where the estimator is $\sqrt{n}$-consistent. Additionally, while the analysis of unlabeled data for extremes presents a significant challenge and is currently a topic of great interest in machine learning for various classification problems, we have developed a semi-supervised framework for the proposed extremile regression using unlabeled data. This framework can also enhance estimation accuracy under model misspecification. Both simulations and real data analyses have been conducted to illustrate the finite sample performance of the proposed methods.
\end{abstract}



\end{frontmatter}
\section{Introduction}
Assessing the extreme behavior of random phenomena is a significant challenge in various fields, including finance, extreme weather and climate events, and medicine \citep{r71}. A common approach to addressing extreme events involves estimating the extreme quantile of a relevant random variable. For instance, this could pertain to the daily return of a stock market index or the intensity of an earthquake. This assumes that the event of interest is represented by a quantitative random variable, denoted as  $\bm{Y}$.
For a fixed $\tau\in (0,1)$,
the $\tau$-th quantile of $\bm{Y}$ is $q_{\tau}$, which is obtained by minimizing the quantile-check function \citep{r5}:
\begin{equation}
	\begin{split}
		q_{\tau}=\arg\min_{\theta\in \mathbb{R}}
		\text{E}\left\{ \left|\tau-\text{I}(\bm{Y}\leq \theta)  \right|\cdot|\bm{Y}-\theta|\right\}.
	\end{split}
\end{equation}
\par
At the same time, expectiles \citep{r70} can serve this purpose as well. While both quantiles and expectiles have been valuable tools, they have faced criticism in the literature for various axiomatic or practical reasons. Quantiles rely solely on whether an observation is below or above a specific value, while expectiles can lack a transparent interpretation due to a lack of explicit expression. 
Additionally, the expected shortfall (ES) has gained popularity as a risk measure that addresses the limitations of quantiles by anticipating losses above the quantile at a certain high level. Regrettably, this approach has led to ES being viewed as a conservative measure of risk, which has made it less appealing to individual financial institutions.

\par
\cite{r1} proposed that the $\tau$-th quantile (1.1) can be derived from the alternative minimization problem
\begin{equation}
	\begin{split}
		q_{\tau}=\arg\min_{\theta\in \mathbb{R}}
		\text{E}\left\{\text{J}_{\tau}(\text{F}(\bm{Y}))\cdot|\bm{Y}-\theta|\right\},
	\end{split}
\end{equation}
where $\text{F}(\cdot)$ is the distribution of $\bm{Y}$, $\text{J}_{\tau}(t)=\partial\text{H}_{\tau}(t)/\partial t$ with
\begin{equation*}
	\begin{split}
		\text{H}_{\tau}(t)=\left \{
		\begin{array}{ll}
			1-(1-t)^{s(\tau)},&\textrm{if}~0< \tau\leq 1/2,\\
			t^{r(\tau)},&\textrm{if}~1/2\leq \tau<1,
		\end{array}
		\right.
	\end{split}
\end{equation*}
being a distribution function with support $[0,1]$ and
$r(\tau)=s(1-\tau)=\log(1/2)/\log(\tau)$.
Based on the equation (1.2), \cite{r1} defined the order-$\tau$ extremile of $\bm{Y}$, which substitutes
the squared deviations in place of the absolute deviations:
\begin{equation}
	\begin{split}
		\xi_{\tau}=\arg\min_{\theta\in \mathbb{R}}
		\text{E}\left\{\text{J}_{\tau}(\text{F}(\bm{Y}))\cdot(\bm{Y}-\theta)^2\right\}.
	\end{split}
\end{equation}
\par
Extremiles, acting as the least squares analogue of quantiles, establish a coherent risk measure determined by weighted expectations, rather than relying on tail probabilities. Therefore, based on the definition of expected extreme value regression (which calculates the expectation of the minimum or maximum tail value) and its effectiveness in measuring risk, this approach can be employed for quantitative risk analysis of extreme events. Examples include earthquakes (such as Turkey's magnitude 7.8 earthquake in 2023), typhoons (like Typhoon Lichma in 2019) and financial crises (such as the U.S. subprime crisis in 2008).
\par
Recently, \cite{r41}  proposed
the conditional order-$\tau$ extremile of $\bm{Y}$ given $\bm{X}=\bm{x}$ as
\begin{equation}
	\begin{split}
		\xi_{\tau}(\bm{x})=\arg\min_{\theta\in \mathbb{R}}
		\text{E}\left\{\text{J}_{\tau}(\text{F}(\bm{Y|\bm{X}}))\cdot(\bm{Y}-\theta)^2|\bm{X}=\bm{x}\right\},
	\end{split}
\end{equation}
where $\text{F}(\cdot|\bm{X})$ is the conditional distribution of $\bm{Y}$ given $\bm{X}$.
\cite{r50} developed estimation methods of extreme conditional extremiles in the framework of heteroscedastic regression model with
heavy-tail noises.
From (1.3) and (1.4), the linear extremile regression can be defined as
\begin{equation}
	\begin{split}
		\xi_{\tau}(\bm{X})=\bm{X}^{\top}\bm{\beta}_{\tau},
	\end{split}
\end{equation}
where
\begin{equation}
	\begin{split}
		\bm{\beta}_{\tau}=\arg\min_{\bm{\beta}\in \mathbb{R}}
		\text{E}\left\{\text{J}_{\tau}(\text{F}(\bm{Y|\bm{X}}))\cdot(\bm{Y}-\bm{X}^{\top}\bm{\beta})^2\right\}.
	\end{split}
\end{equation}
Based on (1.6), we can estimate $\bm{\beta}_{\tau}$ as
\begin{equation}
	\begin{split}
		\bar{\bm{\beta}}_{\tau}=&\arg\min_{\bm{\beta}}\sum_{i=1}^{n}
		\text{J}_{\tau}(\hat{\text{F}}(Y_i|\bm{X}_i))\cdot(Y_i-\bm{X}_i^{\top}\bm{\beta})^2\\
		=&(\bm{X}^{\top}\hat{\bm{W}}\bm{X})^{-1}\bm{X}^{\top}\hat{\bm{W}}\bm{Y},
	\end{split}
\end{equation}
where $\hat{\bm{W}}=\text{diag}\{\text{J}_{\tau}(\hat{\text{F}}(Y_1|\bm{X}_1)),\ldots,
\text{J}_{\tau}(\hat{\text{F}}(Y_n|\bm{X}_n))\}$.
Here we use $\hat{\text{F}}(Y|\bm{X}))$  for an estimator of conditional distribution function $\text{F}(\cdot|\bm{X})$.
If $\text{F}(\cdot|\bm{X})$ is known, $\bar{\bm{\beta}}_{\tau}$ based on $\text{F}(\cdot|\bm{X})$ is $\sqrt{n}$-consistent (see Theorem B.1 for details).

However,  $\text{F}(\cdot|\bm{X})$ is often unknown in practice, then the  widely used  Nadaraya-Watson type estimator of $\text{F}(y|\bm{x})$ is given by
\begin{equation}
	\begin{split}
		\hat{\text{F}}(y|\bm{x})
		=\frac{\sum_{i=1}^{N}\text{I}(Y_i\leq y)\text{K}_h\left(\bm{X}_i-\bm{x}\right)}{\sum_{i=1}^{N}
			\text{K}_h\left(\bm{X}_i-\bm{x}\right)},
	\end{split}
\end{equation}
where $\text{K}_h(\cdot)=\text{K}(\cdot/h)$, $\text{K}(\cdot)$ is a kernel function and $h$ is a bandwidth.

Then, the convergence speed of $\bar{\bm{\beta}}_{\tau}$ in (1.7) with $\hat{\text{F}}(Y_i|\bm{X}_i)$ is $n^{2/5}$, because $\hat{\text{F}}(y|\bm{x})$ is $n^{2/5}$-consistent (see Theorem B.2 for details).
This poses a challenge for parameter estimators, as the preference is for parameter estimation to be preferably $\sqrt{n}$-consistent. To address this, we defined a new linear extremile regression model using a linear quantile regression model. This model exclusively involves unknown parameters, without the inclusion of the unknown non-parametric component as in equation (1.6). Consequently, we are able to establish a $\sqrt{n}$-consistent estimator for the unknown parameters in the linear extremile regression model.
\par
Analyzing unlabeled data for extreme events is inherently challenging. Recently, semi-supervised learning has gained significant traction in extreme machine learning, especially for various classification problems,  see \cite{huang2014semi}, \cite{pei2018robust}, \cite{zhang2023semi} and among others. This is primarily because acquiring labeled data, especially for extreme learning processes, can often be more arduous than obtaining unlabeled data in many applications. In situations where there's an abundance of unlabeled data but only a limited amount of labeled data, there's a strong incentive to leverage the unlabeled data to enhance the predictive performance of a given supervised learning algorithm through the application of suitable semi-supervised (SS) learning methods.
\par
Unlike traditional statistical learning settings, which are typically either supervised or unsupervised, an SS setting represents a convergence of these two realms. It encompasses two distinct data sets: (i) a labeled data set comprising observations for an outcome $\bm{Y}$ and a set of covariates $\bm{X}$, and (ii) a significantly larger unlabeled data set where only $\bm{X}$ is observed. This fundamental distinction sets SS settings apart from standard missing data problems, where the proportion is always assumed to be bounded away from 0, a condition commonly referred to as the positivity (or overlap) assumption in the missing data literature, which is inherently violated in this context.
\par
For example, in biomedical applications, SS settings are assuming growing importance in modern integrative genomics, particularly in the study of expression quantitative trait loci (eQTL) \citep{r59}. This approach merges genetic association studies with gene expression profiles. However, a prevailing challenge in such studies is that due to the limited scale of costly gene expression data, their capabilities are often inadequate  \citep{r61}. Conversely, genetic variation recording is more cost-effective and can typically be used for large-scale datasets, naturally leading to SS settings. Consequently, effective strategies are required to leverage this additional information for generating more potent association mapping methods to detect causal effects of genetic variation. Additionally, SS settings are increasingly relevant in modern studies involving extensive databases, such as machine learning applications in text mining, web page classification, speech recognition, natural language processing, and electronic health records.
\par
Effectively integrating labeled and unlabeled data for comprehensive analysis, to enhance parameter estimation accuracy, is the central concern in SS learning. This area has garnered increasing attention as one of the most promising domains in statistics and machine learning in recent years, encompassing fields like image processing \citep{r56}, anomaly detection \citep{r57}, empirical risk \citep{r55}, and linear regression \citep{r64,r58,r62,r63,r47}. A comprehensive overview of SS learning and recent advancements can be found in \cite{r54} and \cite{r65}.
\par
In this paper, we present a methodology for utilizing unlabeled data to devise SS learning methods for linear extremile regression. This approach not only constructs an effective estimation method when the working model is incorrectly specified but also enhances the estimation accuracy of supervised learning under miss-specification, performing as efficiently as supervised learning when the working model is correctly specified. This further highlights the appeal of the proposed linear extreme regression due to its simplicity and interpretability, even when the working model may only approximate the true model.

\par

To summarize, our statistical contributions are as follows:
\par
(i) We introduce a new definition of linear extremile regression along with an estimation method to ensure the $\sqrt{n}$-consistency of unknown parameters in the model.
\par
(ii) Within a semi-supervised setting, we have developed an estimation method for unknown parameters in the linear extremile regression model using unlabeled data. We have also demonstrated that the resulting estimator proves to be more effective than one utilizing only labeled data.
\par
The remaining sections of this paper are organized as follows: In Section 2, we present the new definition of linear extremile regression and its associated estimation method. Section 3 focuses on the development of semi-supervised learning. Section 4 showcases simulation examples and demonstrates the application of real data to illustrate the proposed methods. Finally, we conclude this paper with a brief discussion in Section 5. All technical proofs are provided in the Appendix.

\section{The new definition of linear extremile regression and its estimation method}
\subsection{The new definition of linear extremile regression}
We now proposed a new estimation method to construct a $\sqrt{n}$-consistent estimator for $\bm{\beta}_{\tau}$ in the linear extremile regression. From Proposition 1 in \cite{r41}, the $\tau$th extremile of $\bm{Y}$ given $\bm{X}$ is
\begin{equation}
	\begin{split}
		\xi_{\tau}(\bm{X})=\int_0^1q_{\bar{\tau}}(\bm{X})\text{J}_{\tau}(\bar{\tau})d\bar{\tau},
	\end{split}
\end{equation}
where $q_{\bar{\tau}}(\bm{X})$ is the conditional $\bar{\tau}$-th quantile of $\bm{Y}$ given $\bm{X}$. The linear quantile regression can be defined as
\begin{equation}
	\begin{split}
		q_{\bar{\tau}}(\bm{X})=\bm{X}^{\top}\bm{\gamma}_{{\bar{\tau}}}.
	\end{split}
\end{equation}
Combine (2.1) and (2.2), we can define the linear extremile regression as
\begin{equation}
	\begin{split}
		\xi_{\tau}(\bm{X})=\int_0^1\bm{X}^{\top}\bm{\gamma}_{\bar{\tau}}
		\text{J}_{\tau}(\bar{\tau})d\bar{\tau}=
		\bm{X}^{\top}\int_0^1\bm{\gamma}_{\bar{\tau}}\text{J}_{\tau}
		(\bar{\tau})d\bar{\tau}\equiv\bm{X}^{\top}\bm{\beta}_{\tau},
	\end{split}
\end{equation}
where
\begin{equation}
	\begin{split}
		\bm{\beta}_{\tau}=\int_0^1\bm{\gamma}_{\bar{\tau}}\text{J}_{\tau}(\bar{\tau})d\bar{\tau}.
	\end{split}
\end{equation}
\par
We now transform the linear extremile regression (1.5) to (2.3) based on the assumption of linear quantile structure (2.2). By comparing (1.6) and (2.4), we can see that only the estimation of unknown parameter $\bm{\gamma}_{\bar{\tau}}$ is present in (2.4), while avoiding the estimation of nonparametric estimation of $\text{F}(\bm{Y|\bm{X}})$ in (1.6).

\begin{pro} Let $\bm{Y}$ given $\bm{X}$ have a finite absolute first moment. Then, for any $\tau\in (0,1)$, we have the following equivalent form expression:
	\begin{equation*}
		\begin{split}
			\xi_{\tau}(\bm{X})=\left \{
			\begin{array}{ll}
				\text{E}\left\{\max(Y_{\bm{X}}^1,\ldots,Y_{\bm{X}}^{r})\right\},&\text{when}~\tau=0.5^{1/r}~\text{with}~r\in \mathbb{N}\backslash \{0\}\\
				\text{E}\left\{\min(Y_{\bm{X}}^1,\ldots,Y_{\bm{X}}^{s})\right\},&\text{when}~\tau=1-0.5^{1/s}~\text{with}~s\in \mathbb{N}\backslash\{0\},
			\end{array}
			\right.
		\end{split}
	\end{equation*}
where $\{Y_{\bm{X}}^i\}$ are independent observations and drawn from the conditional distribution of $\bm{Y}$ given $\bm{X}$.
\end{pro}
\par
From Proposition 2.1, the order-$\tau$ extremile $\xi_{\tau}(\bm{X})$ defined in (2.3) is the same as that in \cite{r41} under the assumption of linear structure. In addition, when $\bm{X}=\bm{1}$, the linear extremile regression (2.3) is equal to extremile in \cite{r1} according to $\xi_{\tau}(\bm{X})=\bm{\beta}_{\tau}=\int_0^1\bm{\gamma}_{\bar{\tau}}\text{J}_{\tau}(\bar{\tau})d\bar{\tau}=\int_0^1q_{\bar{\tau}}\text{J}_{\tau}(\bar{\tau})d\bar{\tau}=\xi_{\tau}$. Therefore, the proposed new definition of linear extremile regression (2.3) is reasonable.

\subsection{Supervised learning}
In this section, we estimate the unknown parameter $\bm{\beta}_{\tau}$ in (2.4) under labeled dataset $\mathcal{L}=\{Y_i,{\bf X}_i\}_{i=1}^n$, which are $n$ independent and identically distributed (i.i.d) observations from $\{\bm{Y},\bm{X}^{\top}\}^{\top}$ in model (2.3).
\par
How to estimate $\bm{\gamma}_{\bar{\tau}}$ in equation (2.4) is difficult, because the ordinary quantile regression (QR) estimation \citep{r5} is a single-point estimation for quantile level $\bar{\tau}$. Therefore, it is hard to obtain the estimator of $\bm{\beta}_{\tau}=\int_0^1\bm{\gamma}_{\bar{\tau}}\text{J}_{\tau}(\bar{\tau})d\bar{\tau}$ according to the integration of quantile function. To solve the above problem,
we consider parameterizing $\bm{\gamma}_{\bar{\tau}}$ into a function  of $\bar{\tau}$ \citep{r44} as
\begin{equation}
	\begin{split}
		\bm{\gamma}_{\bar{\tau}}=\bm{\alpha}_0\bm{b}(\bar{\tau}),
	\end{split}
\end{equation}
where $\bm{\alpha}_0$ is a $p\times q$ unknown matrix and $\bm{b}(\bar{\tau})$ is a set of $q$ known functions of $\bar{\tau}$. Compared with standard QR, the
parameterized method in (2.5) presents numerous advantages, which include simpler estimation and inference according to the smooth objective function (2.7), improved efficiency and better interpretation of the results \citep{r51}.
\par
As mentioned in \cite{r44} and \cite{r45}, valid choices of $\bm{b}(\bar{\tau})$ are, for example, $\bm{b}(\bar{\tau})=\{1,\bar{\tau},\bar{\tau}^2,\bar{\tau}^3\}^{\top}$ simply consists of polynomials of increasing orders, $\bm{b}(\bar{\tau})=\{1,\log(\bar{\tau}),-\log(1-\bar{\tau})\}^{\top}$ relates to an asymmetric logistic distribution and $\bm{b}(\bar{\tau})=\{1,\Phi^{-1}(\bar{\tau}),\sqrt{-2\log(1-\bar{\tau})}\}^{\top}$
is a combination of quantile functions of standard normal $\Phi$ and Rayleigh distributions.
The simulation results of Table 1 in \cite{r49} show that the parametrized method (2.5) is not sensitive
to the selection of $\bm{b}(\bar{\tau})$.
\par
From (2.4) and (2.5), we have
\begin{equation}
	\begin{split}
		\bm{\beta}_{\tau}=\int_0^1\bm{\gamma}_{\bar{\tau}}\text{J}_{\tau}(\bar{\tau})d\bar{\tau}=
		\bm{\alpha}_0\int_0^1\bm{b}(\bar{\tau})\text{J}_{\tau}(\bar{\tau})d\bar{\tau}.
	\end{split}
\end{equation}
Note that the equation (2.6), we only need estimate the unknown parameter $\bm{\alpha}_0$
which is independent of $\bar{\tau}$, instead of estimating the regression coefficients for each quantile $\bar{\tau}\in (0,1)$.
The term
$\int_0^1\bm{b}(\bar{\tau})\text{J}_{\tau}(\bar{\tau})d\bar{\tau}$ is a constant vector under given $\tau$ and
is independent of data $\{\bm{Y},\bm{X}\}$, because $\bm{b}(\cdot)$ and $\text{J}_{\tau}(\cdot)$ are known function. Therefore, for different $\tau$s, just calculate $\int_0^1\bm{b}(\bar{\tau})\text{J}_{\tau}(\bar{\tau})d\bar{\tau}$ that has nothing to do with data, without recalculating $\bm{\alpha}_0$.
In summary, the proposed method is convenient and can be used for big data analysis.
\par
Based on (2.2) and (2.5), we have
 $q_{\bar{\tau}}(\bm{X})=\bm{X}^{\top}\bm{\gamma}_{\bar{\tau}}=\bm{X}^{\top}\bm{\alpha}_0\bm{b}(\bar{\tau})$. Therefore, we can estimate
$\bm{\alpha}_0$ in traditional supervised setting as the minimizer of the integrated objective function:
\begin{equation}
	\begin{split}
		\hat{\bm{\alpha}}=
		\arg\min_{\bm{\alpha}}\sum_{i=1}^{n}L(Y_i,\bm{X}_i,\bm{\alpha})
		=\arg\min_{\bm{\alpha}}\sum_{i=1}^{n}\int_0^1\rho_{\bar{\tau}}(Y_i-\bm{X}_i^{\top}\bm{\alpha}\bm{b}(\bar{\tau}))d\bar{\tau},
	\end{split}
\end{equation}
where  $L(Y_i,\bm{X}_i,\bm{\alpha})=\int_0^1\rho_{\bar{\tau}}(Y_i-\bm{X}_i^{\top}\bm{\alpha}\bm{b}(\bar{\tau}))d\bar{\tau}$ and $\rho_{\bar{\tau}}(r)=\bar{\tau}r-r\text{I}(r<0)$ is the quantile check function.
The objective function $L(Y_i,\bm{X}_i,\bm{\alpha})$ in (2.7)
can be regarded as an average loss function, achieved by marginalizing $
\rho_{\bar{\tau}}(\bm{Y}_i-\bm{X}_i^{\top}\bm{\alpha}\bm{b}(\bar{\tau}))$
over the entire interval $(0,1)$. In addition, the solution of minimizing (2.7) is currently implemented by the
\verb"iqr" function in the \verb"qrcm R" package.
Then, the estimator $\hat{\bm{\beta}}_{\tau}$ with $\hat{\bm{\alpha}}$ in the equation (2.7) is
\begin{equation}
	\begin{split}
		\hat{\bm{\beta}}_{\tau}=
		\hat{\bm{\alpha}}\int_0^1\bm{b}(\bar{\tau})\text{J}_{\tau}(\bar{\tau})d\bar{\tau}.
	\end{split}
\end{equation}
\par
Note that (2.8), it permits estimating the entire extremile  process rather than only obtaining a discrete set of extremiles. If $\int_0^1\bm{b}(\bar{\tau})\text{J}_{\tau}(\bar{\tau})d\bar{\tau}$ in (2.8) is not integrable, we can use
$n^{-1}\sum_{i=1}^n\bm{b}(i/n)\text{J}_{\tau}(i/n)$ to calculate it approximately.
\begin{remark}
	\cite{r43} also consider $\tau$th linear extremile regression as following:
	\begin{equation*}
		\begin{split}
			\hat{\bm{\beta}}_{\tau}=\arg\min_{\bm{\beta}}\sum_{i=1}^{n}
			\text{J}_{\tau}(\hat{F}(\bm{Y}_i|\bm{X}))(\bm{Y}_i-\bm{X}_i^{\top}
			\bm{\beta})^2.
		\end{split}
	\end{equation*}
	Note that $\hat{F}(\bm{Y}_i|\bm{X})$ is wrong in above equation.
	For their assumed conditions and theorem results, $\hat{F}(\bm{Y}_i|\bm{X})$ should be $\text{F}(\bm{Y}_i|\bm{X}_i)$. If is $\hat{F}(\bm{Y}_i|\bm{X}_i)$, $\|\hat{\bm{\beta}}_{\tau}-\bm{\beta}_{\tau}\|_2=O_p\left(\sqrt{\frac{\log p(n+p)}{n+p}}\right)$ is not available.
	And the condition C3 in their paper $\lambda_{\max}\left(\frac{\bm{X}^{\top}\bm{W}\bm{X}+\lambda_2I}{(1+\lambda_2)(n+p)}\right)<\infty$ is wrong with $\bm{W}=\text{diag}\left\{\text{J}_{\tau}(\hat{F}(Y_1|\bm{X}))/n,\ldots,\text{J}_{\tau}(\hat{F}(Y_n|\bm{X})/n)\right\}$, because $\hat{F}(Y_i|\bm{X})$ is a random variable.
\end{remark}

\subsection{Large sample properties}
The asymptotic theories are derived by applying standard asymptotic results \citep{r46}. To facilitate the presentation, we need to introduce some notations.
Let $\text{Vec}(\cdot)$ be the vectoring operation, which creates a column vector by stacking the column vectors of below one another, that is,
$\text{Vec}(\bm{\alpha})=(\bm{\alpha}_1^{\top},\dots,\bm{\alpha}_q^{\top})^{\top}$
with $\bm{\alpha}_j=(\alpha_{1,j},\dots,\alpha_{p,j})^{\top}$.
Denote $S(\bm{\alpha})=n^{-1}\sum_{i=1}^{n}\nabla_{\text{Vec}(\bm{\alpha})}L(Y_i,\bm{X}_i,\bm{\alpha})
=n^{-1}\sum_{i=1}^{n}\int_0^1\bm{b}(\bar{\tau})\otimes\bm{X}_i[\text{I}(Y_i<\{\bm{b}(\bar{\tau})\otimes\bm{X}_i\}^{\top}\text{Vec}(\bm{\alpha}))-\bar{\tau}]d\bar{\tau}$,
where $\otimes$
is the Kronecker product.
\par
{\bf C1}: The true unknown parameter vector $\bm{\alpha}_0$ in (2.6) is an interior point of
a compact set $\Theta$ and
satisfies $E\left\{S(\bm{\alpha}_0)|\bm{X}\right\}={\bf 0}$.
\par
{\bf C2}: The loss function $L(\bm{Y},\bm{X},\bm{\alpha})$ satisfies $E[\sup_{\bm{\alpha}\in \Theta}\{L(\bm{Y},\bm{X},\bm{\alpha})\}^2]<\infty$. $S(\bm{\alpha})$ is continuously differentiable, $E\{\sup_{\bm{\alpha}\in \Theta}\|S(\bm{\alpha})\|_2^2\}<\infty$ and
$E\{\sup_{\bm{\alpha}\in \Theta}\|\nabla_{\text{Vec}(\bm{\alpha})}S(\bm{\alpha})\|\}<\infty$
with $\|\cdot\|$ is the spectral norm.
\par
{\bf C3}: ${\bf H}=E\{\nabla_{\text{Vec}(\bm{\alpha})}S(\bm{\alpha})|_{\bm{\alpha}=\bm{\alpha}_0}\}=
E[\int_0^1\{\bm{X}^{\top}\bm{\alpha}_0\nabla_{\bar{\tau}}\bm{b}(\bar{\tau})\}^{-1}
\{\bm{b}(\bar{\tau})\otimes\bm{X}\}\{\bm{b}(\bar{\tau})\otimes\bm{X}\}^{\top}d\bar{\tau}]
$ is nonsingular.

\begin{remark}
	The validity of the conditions {\bf C1} and {\bf C2} depends on the structure of
	$\bm{b}(\bar{\tau})$, which should induce a well-defined quantile function  such that $\Theta$ is not empty (conditions {\bf C1}); $\bm{b}(\bar{\tau})$ is continuous
	and ensure that a central limit theorem can be applied to $S(\bm{\alpha})$ (conditions {\bf C2}).	
	Conditions {\bf C1} and {\bf C2} are also used in \cite{r44}. Valid choices of $\bm{b}(\bar{\tau})$ are, for example, functions of the form $\bar{\tau}^c$, $\log(\bar{\tau})$, $\log(1-\bar{\tau})$, $c^{\bar{\tau}}$, the quantile function of any distribution with finite moments, splines or a combination of the above.	
	Condition {\bf C3} is needed to establish the asymptotic normality.
\end{remark}
\begin{thm}
	Suppose that the conditions {\bf C1} and {\bf C2} hold. Then as $n\rightarrow\infty$, we have
	\begin{equation*}
		\begin{split}
			\text{Vec}(\hat{\bm{\alpha}}-\bm{\alpha}_0)\xrightarrow{p} \bm{0},
		\end{split}
	\end{equation*}
	where $\xrightarrow{p}$ represents the convergence in the probability.
	Moreover, if condition {\bf C3} holds, we can obtain
	\begin{equation*}
		\begin{split}
			\sqrt{n}\text{Vec}(\hat{\bm{\alpha}}-\bm{\alpha}_0)
			\xrightarrow{L}\textrm{N}\left(\bm{0},{\bf H}^{-1}{\bm \Sigma}{\bf H}^{-1}\right),
		\end{split}
	\end{equation*}
	where $\xrightarrow{L}$ represents the convergence in the distribution and ${\bm \Sigma}=E\{S(\bm{\alpha}_0)S(\bm{\alpha}_0)^{\top}\}$.
\end{thm}
Theorem 2.1 shows that the parameter $\bm{\alpha}_0$ is identified and its estimator
$\hat{\bm{\alpha}}$ has a normal distribution in large samples. The large sample distribution of the plugin estimator of
$\bm{\beta}_{\tau}=\bm{\alpha}_0\int_0^1\bm{b}(\bar{\tau})\text{J}_{\tau}(\bar{\tau})d\bar{\tau}$ can also be obtained as
\begin{equation}
	\begin{split}
		\hat{\bm{\beta}}_{\tau}=
		\hat{\bm{\alpha}}\int_0^1\bm{b}(\bar{\tau})\text{J}_{\tau}(\bar{\tau})d\bar{\tau}=\tilde{\bm{b}}(\tau)^{\top}\text{Vec}(\tilde{\bm{\alpha}}),
	\end{split}
\end{equation}
where $\tilde{\bm{b}}(\tau)=\int_0^1\bm{b}(\bar{\tau})\text{J}_{\tau}(\bar{\tau})d\bar{\tau}\otimes \bm{I}_p$ and $\bm{I}_p$ is the identity matrix of size $p$. Then, we consider the large sample distribution of $\hat{\bm{\beta}}$ by (2.9) in the following theorem.

\begin{thm}
	Suppose that the conditions {\bf C1}-{\bf C3} hold, we have
	\begin{equation*}
		\begin{split}
			\sqrt{n}(\hat{\bm{\beta}}_{\tau}-\bm{\beta}_{\tau})
			\xrightarrow{L}\textrm{N}\left(\bm{0},\tilde{\bm{b}}(\tau)^{\top}{\bf H}^{-1}{\bm \Sigma}{\bf H}^{-1}\tilde{\bm{b}}(\tau)\right).
		\end{split}
	\end{equation*}
\end{thm}

\section{Semi-supervised learning}
\subsection{Data representation}
Let $\mathbb{P}$ be the joint distribution of $\{\bm{Y},\bm{X}^{\top}\}^{\top}$ and let $\mathbb{P}_{X}$ be the marginal distribution of $\bm{X}$.
In semi-supervised setting, the data available are $\mathcal{D}=\mathcal{L}\cup \mathcal{M}$,
where $\mathcal{L}=\{Y_i,{\bf X}_i\}_{i=1}^n$ is from $\mathbb{P}$ and $\mathcal{M}=\{{\bf X}_i\}_{i=n+1}^N$ with $N\geq 1$ are $N$ i.i.d observations from $\mathbb{P}_{X}$. The $n/N\rightarrow\rho$ for some constant $\rho\in[0,+\infty)$ as $n\rightarrow\infty$ and $N\rightarrow\infty$. Note that the
semi-supervised setting allows $n/N\rightarrow0$, that means that the unlabeled dataset can be of much larger size than the labeled one in various practical problems, as labeling of the outcomes is often very costly. However, the missing completely at random assumption is that $\rho>0$, which is the major difference between semi-supervised setting and missing data.
\subsection{The target parameter}
Based on equations (2.3) and (2.6), the new definition $\tau$th linear extremile regression is as follows:
\begin{equation}
	\begin{split}
		\xi_{\tau}(\bm{X})=\bm{X}^{\top}\bm{\alpha}_0\int_0^1\bm{b}(\bar{\tau})\text{J}_{\tau}(\bar{\tau})d\bar{\tau},
	\end{split}
\end{equation}
where $\bm{\alpha}_0$ is the unknown parameter.
\par
In some actual data analysis, the model (3.1) may not be correct because the linear structure assumptions of the model (2.2) or (2.3) may not be correct.
But due to the simplicity and interpretability of the linear structure, it is often continued to be used. Therefore, consider a $\tau$th linear extremile working regression model
$\xi_{\tau}(\bm{X})=\bm{X}^{\top}\bm{\alpha}^{*}\int_0^1\bm{b}(\bar{\tau})\text{J}_{\tau}(\bar{\tau})d\bar{\tau}$,
where the unknown parameter $\bm{\alpha}^{*}$ is defined as
\begin{equation}
	\begin{split}
		\bm{\alpha}^{*}=&\arg\min_{\bm{\alpha}}E\{L(\bm{Y},\bm{X},\bm{\alpha})\}.
	\end{split}
\end{equation}
\par
It is noteworthy that in supervised framework, $\bm{\alpha}^{*}$ is equal to $\bm{\alpha}_0$ when the outcome variable $\bm{Y}$ is fully observed and the working model is correctly specified.
\subsection{Semi-supervised learning}
From the idea in \cite{r47}, let random vector $\bm{Z}\in \mathbb{R}^d$ be a deterministic function of $\bm{X}$ for
some positive integer $d$. Without loss of generality, we fix the first element of $\bm{Z}$ to be 1. With certain
proper choice of $\bm{Z}$, we have
\begin{equation*}
	\begin{split}
		E\{L(\bm{Y},\bm{X},\bm{\alpha})\}=E\{\bm{Z}^{\top}\varphi(\bm{\alpha})\},
	\end{split}
\end{equation*}
where $\varphi(\bm{\alpha})=\{E(\bm{Z}\bm{Z}^{\top})\}^{-1}E\{\bm{Z}L(\bm{Y},\bm{X},\bm{\alpha})\}$.
Then, we propose a new class of loss functions by incorporating the
information from the unlabeled data into the supervised estimation:
\begin{equation}
	\begin{split}
		\tilde{\bm{\alpha}}=&\arg\min_{\bm{\alpha}}\left\{
		\sum_{i=1}^{n}L(Y_i,\bm{X}_i,\bm{\alpha})
		+\sum_{i=n+1}^{n+N}\bm{Z}_i^{\top}\hat{\varphi}(\bm{\alpha})\right\}\\
		=&\arg\min_{\bm{\alpha}}\sum_{i=1}^{n}\omega_iL(Y_i,\bm{X}_i,\bm{\alpha}),
	\end{split}
\end{equation}
where
$\hat{\varphi}(\bm{\alpha})=\left(\sum_{i=1}^{n}\bm{Z}_i\bm{Z}_i^{\top}/n\right)^{-1}
\sum_{i=1}^{n}\bm{Z}_iL(Y_i,\bm{X}_i,\bm{\alpha})/n$
and
$$\omega_i=1+N/n\left(\sum_{i=n+1}^{n+N}\bm{Z}_i/N\right)^{\top}
\left(\sum_{i=1}^{n}\bm{Z}_i\bm{Z}_i^{\top}/n\right)^{-1}\bm{Z}_i.$$
\par
We can select $\bm{Z}$ by $\text{GBIC}_{ppo}$ as suggestion in \cite{r47}.
In addition, the solution of minimizing (3.3) is currently implemented by the
\verb"iqr" function in the \verb"qrcm R" package. From the equations (2.6) and (3.3), we can obtain
\begin{equation}
	\begin{split}
		\tilde{\bm{\beta}}_{\tau}=
		\tilde{\bm{\alpha}}\int_0^1\bm{b}(\bar{\tau})\text{J}_{\tau}(\bar{\tau})d\bar{\tau}.
	\end{split}
\end{equation}

\subsection{Large sample properties}
The following conditions are needed to establish the consistency and asymptotic normality of $\tilde{\bm{\alpha}}$.
\par
{\bf C4}: The unknown parameter vector $\bm{\alpha}^{*}$ in (3.2) is an interior point of
a compact set $\Theta$,
and $\tilde{{\bf H}}=E\{\nabla_{\text{Vec}(\bm{\alpha})}S(\bm{\alpha})|_{\bm{\alpha}=\bm{\alpha}^{*}}\}
$ is nonsingular.
\par
{\bf C5}: The random vector $\bm{Z}$ is bounded almost surely and ${\bm \Sigma}_{\bm{Z}}=E(\bm{Z}\bm{Z}^{\top})$
is nonsingular.
\begin{remark}
	The condition {\bf C4} ensures the uniqueness of $\bm{\alpha}^{*}$ under the strict convexity of $E\{L(\bm{Y},\bm{X},\bm{\alpha})\}$.
	Condition {\bf C5} is regularity condition for the unlabeled data. Conditions {\bf C4} and {\bf C5} are also used in \cite{r47}.
\end{remark}

\begin{thm}
	Suppose that the conditions {\bf C2}, {\bf C4} and {\bf C5} hold and $n/N\rightarrow\rho$ for some constant $\rho\in[0,+\infty)$ as $n\rightarrow\infty$ and $N\rightarrow\infty$. Then as $n\rightarrow\infty$ and $N\rightarrow\infty$, we have
	\begin{equation*}
		\begin{split}
			\text{Vec}(\tilde{\bm{\alpha}}-\bm{\alpha}^{*})\xrightarrow{p} \bm{0},
		\end{split}
	\end{equation*}
	and
	\begin{equation*}
		\begin{split}
			\sqrt{n}\text{Vec}(\tilde{\bm{\alpha}}-\bm{\alpha}^{*})
			\xrightarrow{L}\textrm{N}\left(\bm{0},\tilde{{\bf H}}^{-1}{\bm \Sigma}_{\rho}\tilde{{\bf H}}^{-1}\right),
		\end{split}
	\end{equation*}
	where ${\bm \Sigma}_{\rho}=E(\bm{W}\bm{W}^{\top})+\rho E(\bm{V}\bm{V}^{\top})$,
	$\bm{W}=S(\bm{\alpha}^{*})-N(n+N)^{-1}\bm{A}^{\top}\bm{Z}$,
	$\bm{V}=N(n+N)^{-1}\bm{A}^{\top}\bm{Z}$ and $\bm{A}={\bm \Sigma}_{\bm{Z}}^{-1}E\{\bm{Z}S(\bm{\alpha}^{*})^{\top}\}$.
\end{thm}

\begin{thm}
	Suppose that conditions in Theorem 3.1 hold, we have
	\begin{equation*}
		\begin{split}
			\sqrt{n}(\tilde{\bm{\beta}}_{\tau}-\bm{\beta}^{*}_{\tau})
			\xrightarrow{L}\textrm{N}\left(\bm{0},\tilde{\bm{b}}(\tau)^{\top}\tilde{{\bf H}}^{-1}{\bm \Sigma}_{\rho}\tilde{{\bf H}}^{-1}\tilde{\bm{b}}(\tau)\right),
		\end{split}
	\end{equation*}
	where $\tilde{\beta}_{\tau}=\tilde{\alpha}\int_0^1\bm{b}(\bar{\tau})\text{J}_{\tau}(\bar{\tau})d\bar{\tau}$ and
	$\bm{\beta}^{*}_{\tau}=\bm{\alpha}^{*}\int_0^1\bm{b}(\bar{\tau})\text{J}_{\tau}(\bar{\tau})d\bar{\tau}$
	.
\end{thm}
\subsection{Comparison between supervised learning and semi-supervised learning}
Theorems 2.2 and 3.2 have important implication on the asymptotic efficiency comparison between the supervised estimator $\hat{\bm{\beta}}_{\tau}$ and the semi-supervised estimator $  \tilde{\bm{\beta}}_{\tau}$.
From Theorem 2.2, ${\bm \Sigma}$ can be write as the following form as
$\bm{U}$ and $\bm{A}^{\top}\bm{Z}$ are uncorrelated,
$${\bm \Sigma}=E(\bm{U}\bm{U}^{\top})+E\left\{(\bm{A}^{\top}\bm{Z})
(\bm{A}^{\top}\bm{Z})^{\top}\right\},$$ where $\bm{U}=S(\bm{\alpha}^{*} )-\bm{A}^{\top}\bm{Z}$. Moreover, ${\bm \Sigma}_{\rho}$
can be rewrite as:
$${\bm \Sigma}_{\rho}=E(\bm{U}\bm{U}^{\top})+\frac{n}{n+N}E\left\{(\bm{A}^{\top}\bm{Z})(\bm{A}^{\top}\bm{Z})^{\top}\right\}.$$
Note that $n/(n+N)\leq 1$.
Therefore, the semi-supervised estimator $\tilde{\bm{\beta}}_{\tau}$
is equally or more efficient than the supervised estimator $\hat{\bm{\beta}}_{\tau}$
according to ${\bm \Sigma}_{\rho}\leq{\bm \Sigma}$.
\subsection{Estimation of covariance}
Finally, we provide consistent analytical estimators of the components of the variances below:
\begin{equation*}
	\begin{split}
		&\hat{{\bf H}}(\bm{\alpha})=\frac{1}{n}\sum_{i=1}^{n}
		\int_0^1\{\bm{X}_i^{\top}\bm{\alpha}\nabla_{\bar{\tau}}\bm{b}(\bar{\tau})\}^{-1}
		\{\bm{b}(\bar{\tau})\otimes\bm{X}_i\}\{\bm{b}(\bar{\tau})\otimes\bm{X}_i\}^{\top}d\bar{\tau},\\
		&\hat{{\bm \Sigma}}_{\rho}= \frac{1}{n}\sum_{i=1}^{n}\hat{\bm{W}}_i\hat{\bm{W}}_i^{\top}+\frac{1}{N}\sum_{i=n+1}^{n+N}\hat{\bm{V}}_i\hat{\bm{V}}_i^{\top},\\
		&\hat{{\bm \Sigma}}=\frac{1}{n}\sum_{i=1}^{n}\hat{S}_i(\hat{\bm{\alpha}})\hat{S}_i(\hat{\bm{\alpha}})^{\top},
	\end{split}
\end{equation*}
where $\hat{\bm{W}}_i=\hat{S}_i(\tilde{\bm{\alpha}})-N(n+N)^{-1}\hat{\bm{A}}^{\top}\bm{Z}_i,$
$\hat{\bm{V}}_i=N(n+N)^{-1}\hat{\bm{A}}^{\top}\bm{Z}_i,$
$\hat{S}_i(\bm{\alpha})=\bm{b}(\bar{\tau})\otimes\bm{X}_i\int_0^1[\text{I}(Y_i<\{\bm{b}(\bar{\tau})\otimes\bm{X}_i\}^{\top}\text{Vec}(\bm{\alpha}))-\bar{\tau}]d\bar{\tau}$
and
$\hat{\bm{A}}=(\sum_{i=1}^{n}\bm{Z}_i\bm{Z}_i)^{-1}\sum_{i=1}^{n}\{\bm{Z}_i\hat{S}_i(\tilde{\bm{\alpha}})^{\top}\}$.
The consistency of these estimators follows from the law of large numbers and consistency of $\text{Vec}(\hat{\bm{\alpha}})$ and $\text{Vec}(\tilde{\bm{\alpha}})$, as all the components are continuous functions of the parameters. Therefore, the limiting covariance matrices $\tilde{\bm{b}}(\tau)^{\top}{\bf H}^{-1}{\bm \Sigma}{\bf H}^{-1}\tilde{\bm{b}}(\tau)$ and $\tilde{\bm{b}}(\tau)^{\top}\tilde{{\bf H}}^{-1}{\bm \Sigma}_{\rho}\tilde{{\bf H}}^{-1}\tilde{\bm{b}}(\tau)$ in Theorems 2.2 and 3.2
can be estimated by $\tilde{\bm{b}}(\tau)^{\top}\{\hat{{\bf H}}(\hat{\bm{\alpha}})\}^{-1}\hat{{\bm \Sigma}}_{\rho}\{\hat{{\bf H}}(\hat{\bm{\alpha}})\}^{-1}\tilde{\bm{b}}(\tau)$ and $\tilde{\bm{b}}(\tau)^{\top}\{\hat{{\bf H}}(\tilde{\bm{\alpha}})\}^{-1}\hat{{\bm \Sigma}}\{\hat{{\bf H}}(\tilde{\bm{\alpha}})\}^{-1}\tilde{\bm{b}}(\tau)$, respectively.

\section{Numerical studies}
In this section, we first use Monte Carlo simulation studies to assess the finite sample performance of the proposed procedures and then demonstrate the application of the proposed methods with a real data analysis. All programs are written in \textsf{R} code.
\subsection{Simulation example 1: the performance of supervised learning}
In this section, we study the performances of the supervised learning in section 2. We compare the supervised learning (SL) $\hat{\bm{\beta}}_{\tau}$ in (2.8) with ordinary estimator (OE) $\bar{\bm{\beta}}_{\tau}$ in (1.7).
We generate data from the following linear model:
\begin{equation}
	\begin{split}
		{\bf Y}={\bf X}^{\top}{\bf\bm{\beta}}_0+\sigma({\bf X})(\bm{\varepsilon}-\hat{e}_{\tau}),
	\end{split}
\end{equation}
where ${\bf X}=({\bf 1},{\bf X}_{1},{\bf X}_{2})^{\top}$, ${\bf X}_{1}$ and ${\bf X}_{2}$ are drawn from a uniform distribution $U(0,1)$.
The true value of the parameter is ${\bf \bm{\beta}}_0=(1,2,3)^{\top}$.
The $
\hat{e}_{\tau}=\sum_{i=1}^{\tilde{n}}
\left\{\text{H}_{\tau}\left(\frac{i}{\tilde{n}}\right)-\text{H}_{\tau}\left(\frac{i-1}{\tilde{n}}\right)\right\}\varepsilon_{i,\tilde{n}}
$ is the estimator of the $\tau$-level extremile of $\bm{\varepsilon}$ \citep{r1}, which is used to eliminate the influence of different $\tau$-level extremile,
where $\varepsilon_{1,\tilde{n}}\leq \varepsilon_{2,\tilde{n}}\leq\cdots\leq \varepsilon_{\tilde{n},\tilde{n}}$ denotes the ordered sample and $\tilde{n}=10^6$.
Therefore, we have $\xi_{\tau}(\bm{X})=\bm{X}^{\top}\bm{\beta}_{0}$ by the setting of model (4.1).
Three error distributions of $\bm{\varepsilon}$ are considered:
a standard normal distribution $N(0,1)$, a t distribution with 5 degrees of freedom $t(5)$ and a uniform distribution $U(0,1)$. Two case of $\sigma({\bf X})$ are considered:
$\sigma({\bf X})=0.5$ and $\sigma({\bf X})=0.4\sqrt{1+|{\bf X}_{1}|+|{\bf X}_{2}|}$.
\par
To evaluate the performance of the estimation method, we calculate the total absolute error : $\text{TAE}=\sum_{j=1}^3|{\bf \hat{\bm{\beta}}}_{\tau,j}-{\bf \bm{\beta}}_{0,j}|$.
The simulation results of the means of TAEs based on $\tau=0.1,0.3,0.5,0.7,0.9$ are
shown in Tables 1 and 2, which are based on 500 simulation replications and the sample size is $n=500$.
The simulation results in Tables 1 and 2 show that the performances of SL are better than those of OE under different errors, $\tau$s and $\sigma({\bf X})$. This means that our proposed method SL
has indeed improved the estimation accuracy compared to OE.
\begin{table}[htp]
	\footnotesize
	\caption{The means and standard deviations (in parentheses) of TAEs with different errors, $\tau$s and methods under $\sigma({\bf X})=0.5$.  }
	\centering
	\begin{tabular}{@{}c|c|ccccc@{}}
		\hline
		Error&Method&$\tau=0.1$&$\tau=0.3$&$\tau=0.5$&$\tau=0.7$&$\tau=0.9$\\
		\hline
		N(0,1)&OE& 0.526~(0.220)& 0.295~(0.135)& 0.179~(0.093)& 0.224~(0.112)& 0.525~(0.232)\\
		&SL& 0.250~(0.131)& 0.188~(0.101)& 0.178~(0.092)& 0.188~(0.098)& 0.259~(0.136)\\
		\hline
		t(5)&OE& 0.759~(0.369)& 0.340~(0.169)& 0.230~(0.120)& 0.295~(0.158)& 0.808~(0.388)\\
		&SL& 0.384~(0.179)& 0.237~(0.120)& 0.222~(0.113)& 0.242~(0.125)& 0.388~(0.184)\\
		\hline
		U(0,1)&OE& 0.203~(0.079)& 0.113~(0.039)& 0.051~(0.026)& 0.080~(0.020)& 0.131~(0.055)\\
		&SL& 0.044~(0.034)& 0.051~(0.028)& 0.051~(0.026)& 0.052~(0.028)& 0.042~(0.023)\\
		\hline
	\end{tabular}
\end{table}

\begin{table}[htp]
	\footnotesize
	\caption{The means and standard deviations (in parentheses) of TAEs with different errors, $\tau$s and methods under $\sigma({\bf X})=0.4\sqrt{1+|{\bf X}_{1}|+|{\bf X}_{2}|}$.  }
	\centering
	\begin{tabular}{@{}c|c|ccccc@{}}
		\hline
		Error&Method&$\tau=0.1$&$\tau=0.3$&$\tau=0.5$&$\tau=0.7$&$\tau=0.9$\\
		\hline
		N(0,1)&OE& 0.520~(0.221)& 0.330~(0.155)& 0.182~(0.094)& 0.243~(0.128)& 0.513~(0.256)\\
		&SL& 0.275~(0.140)& 0.199~(0.106)& 0.181~(0.093)& 0.202~(0.104)& 0.284~(0.143)\\
		\hline
		t(5)&OE& 0.808~(0.442)& 0.371~(0.194)& 0.249~(0.134)& 0.308~(0.163)& 0.815~(0.427)\\
		&SL& 0.413~(0.203)& 0.244~(0.133)& 0.234~(0.124)& 0.264~(0.130)& 0.426~(0.203)\\
		\hline
		U(0,1)&OE& 0.195~(0.065)& 0.112~(0.039)& 0.056~(0.031)& 0.089~(0.022)& 0.152~(0.073)\\
		&SL& 0.047~(0.030)& 0.056~(0.030)& 0.055~(0.031)& 0.056~(0.028)& 0.044~(0.022)\\
		\hline
	\end{tabular}
\end{table}

\subsection{Simulation example 2: the performance of semi-supervised learning}
In this section, we study the performances of the semi-supervised learning in section 3. We compare the semi-supervised learning $\tilde{\bm{\beta}}_{\tau}$ in (3.4) with the supervised learning $\hat{\bm{\beta}}_{\tau}$ in (2.8).
We generate data from the following model in \cite{r47}:
\begin{equation*}
	\begin{split}
		{\bf Y}=\alpha_0+{\bf X}^{\top}{\bf\bm{\alpha}}_1+\alpha_2\sum_{j,k}{\bf X}_{j}{\bf X}_{k}+(1+{\bf X}^{\top}{\bf\bm{\alpha}}_3)\bm{\varepsilon},
	\end{split}
\end{equation*}
where ${\bf X}=({\bf X}_{1},\dots,{\bf X}_{4})^{\top}$, $\{{\bf X}_{j}\}_{j=1}^4$ are drawn from standard normal distribution, and the true value of the parameter is $(\alpha_0,{\bf\bm{\alpha}}_1^{\top},\alpha_2,{\bf\bm{\alpha}}_3^{\top})^{\top}=(1,0.5,0.5,0.5,0.5,1,0.5,0.5,0,0)^{\top}$. Three error distributions of $\bm{\varepsilon}$ are considered:
$N(0,1)$, $t(5)$ and $U(0,1)$. The labeled $(n)$ and three unlabeled $(N)$ sample size are considered:
$n=500$ and $N=500,1000,2000$.
\par
To compare the semi-supervised learning (SSL) with the supervised learning (SL), we report the
estimated Asymptotic Relative Efficiency (ARE), which is defined by the ratio of the empirical standard deviation of estimator of SL and that of SSL.
The simulation results of the means of estimators, their empirical standard deviations and AREs based on $\tau=0.1,0.3,0.5,0.7,0.9$ are shown in Tables 3-5, which are based on 500 simulation replications.
Form the results in Tables 3-5, we can see that the improvements of semi-supervised learning (SSL)
over the supervised learning (SL) are more significant when $N$ gets large according to ARE, which is reasonable.

\begin{table}[htp]
	\footnotesize
	\caption{The AREs and means and standard deviations (in parentheses) of SL and SSL with different $\tau$s and Ns under $\bm{\varepsilon}\sim N(0,1)$.  }
	\centering
	\begin{tabular}{@{}c|c|ccc|cccc@{}}
		\hline
		&&&SSL&&&ARE\\
		\cline{3-8}
		$\tau$&SL&N=500&N=1000&N=2000&N=500&N=1000&N=2000\\
		\hline
		0.1 & 1.772~(0.115) & 1.783~(0.112)& 1.785~(0.110)& 1.786~(0.109)& 1.026& 1.043& 1.059\\
		& 0.054~(0.197) & 0.058~(0.193)& 0.059~(0.188)& 0.057~(0.187)& 1.025& 1.047& 1.053\\
		& 0.070~(0.195) & 0.073~(0.181)& 0.070~(0.180)& 0.073~(0.174)& 1.074& 1.082& 1.121\\
		& 0.491~(0.196) & 0.492~(0.186)& 0.491~(0.185)& 0.493~(0.176)& 1.051& 1.055& 1.111\\
		& 0.508~(0.212) & 0.505~(0.200)& 0.508~(0.192)& 0.502~(0.188)& 1.062& 1.103& 1.127\\
		\hline
		0.3 & 3.420~(0.109) & 3.436~(0.094)& 3.438~(0.087)& 3.439~(0.084)& 1.158& 1.246& 1.289\\
		& 0.304~(0.159) & 0.297~(0.140)& 0.301~(0.138)& 0.295~(0.132)& 1.134& 1.152& 1.198\\
		& 0.295~(0.152) & 0.300~(0.138)& 0.292~(0.132)& 0.295~(0.130)& 1.104& 1.153& 1.173\\
		& 0.507~(0.155) & 0.502~(0.139)& 0.504~(0.136)& 0.503~(0.125)& 1.116& 1.141& 1.244\\
		& 0.501~(0.161) & 0.504~(0.149)& 0.507~(0.138)& 0.507~(0.133)& 1.082& 1.165& 1.207\\
		\hline
		0.5 & 4.907~(0.133) & 4.925~(0.102)& 4.931~(0.092)& 4.931~(0.084)& 1.310& 1.455& 1.591\\
		& 0.391~(0.162) & 0.392~(0.137)& 0.395~(0.136)& 0.393~(0.128)& 1.182& 1.191& 1.263\\
		& 0.393~(0.158) & 0.393~(0.138)& 0.393~(0.134)& 0.398~(0.127)& 1.139& 1.178& 1.237\\
		& 0.502~(0.157) & 0.503~(0.134)& 0.503~(0.133)& 0.501~(0.124)& 1.169& 1.180& 1.260\\
		& 0.501~(0.152) & 0.497~(0.135)& 0.496~(0.124)& 0.501~(0.122)& 1.128& 1.233& 1.250\\
		\hline
		0.7 & 6.419~(0.187) & 6.427~(0.150)& 6.425~(0.126)& 6.430~(0.109)& 1.247& 1.481& 1.711\\
		& 0.508~(0.185) & 0.515~(0.160)& 0.510~(0.150)& 0.512~(0.141)& 1.154& 1.227& 1.307\\
		& 0.499~(0.190) & 0.509~(0.164)& 0.507~(0.151)& 0.507~(0.139)& 1.160& 1.263& 1.364\\
		& 0.496~(0.174) & 0.496~(0.150)& 0.490~(0.147)& 0.493~(0.138)& 1.159& 1.184& 1.259\\
		& 0.502~(0.195) & 0.501~(0.160)& 0.510~(0.151)& 0.500~(0.144)& 1.215& 1.286& 1.349\\
		\hline
		0.9 & 8.929~(0.295) & 8.945~(0.260)& 8.957~(0.255)& 8.954~(0.251)& 1.135& 1.157& 1.175\\
		& 0.581~(0.252) & 0.587~(0.227)& 0.582~(0.212)& 0.586~(0.203)& 1.110& 1.187& 1.241\\
		& 0.607~(0.260) & 0.605~(0.241)& 0.605~(0.234)& 0.600~(0.229)& 1.078& 1.108& 1.133\\
		& 0.504~(0.244) & 0.508~(0.220)& 0.500~(0.218)& 0.501~(0.214)& 1.105& 1.119& 1.139\\
		& 0.514~(0.260) & 0.510~(0.238)& 0.511~(0.214)& 0.513~(0.206)& 1.092& 1.215& 1.260\\
		\hline
	\end{tabular}
\end{table}

\begin{table}[htp]
	\footnotesize
	\caption{The AREs and means and standard deviations (in parentheses) of SL and SSL with different $\tau$s and Ns under $\bm{\varepsilon}\sim t(5)$.  }
	\centering
	\begin{tabular}{@{}c|c|ccc|cccc@{}}
		\hline
		&&&SSL&&&ARE\\
		\cline{3-8}
		$\tau$&SL&N=500&N=1000&N=2000&N=500&N=1000&N=2000\\
		\hline
		0.1 &~1.611~(0.129) &~1.625~(0.125)&~1.631~(0.125)&~1.634~(0.123)& 1.030& 1.032& 1.049\\
		&-0.064~(0.222) &-0.070~(0.205)&-0.067~(0.204)&-0.074~(0.201)& 1.083& 1.088& 1.104\\
		&-0.090~(0.210) &-0.083~(0.204)&-0.085~(0.199)&-0.081~(0.198)& 1.032& 1.059& 1.061\\
		&~0.498~(0.216) &~0.498~(0.209)&~0.493~(0.203)&~0.494~(0.202)& 1.031& 1.061& 1.069\\
		&~0.503~(0.222) &~0.500~(0.212)&~0.502~(0.206)&~0.500~(0.204)& 1.048& 1.078& 1.093\\
		\hline
		0.3 & 3.370~(0.117) & 3.378~(0.102)& 3.383~(0.094)& 3.385~(0.090)& 1.143& 1.250& 1.306\\
		& 0.222~(0.157) & 0.218~(0.140)& 0.216~(0.139)& 0.219~(0.137)& 1.122& 1.128& 1.141\\
		& 0.229~(0.168) & 0.225~(0.147)& 0.223~(0.143)& 0.221~(0.136)& 1.148& 1.180& 1.242\\
		& 0.494~(0.174) & 0.493~(0.152)& 0.493~(0.151)& 0.494~(0.144)& 1.148& 1.146& 1.209\\
		& 0.501~(0.174) & 0.504~(0.160)& 0.501~(0.149)& 0.505~(0.143)& 1.084& 1.166& 1.217\\
		\hline
		0.5 &~4.922~(0.140) &~4.926~(0.109)&~4.932~(0.102)&~4.937~(0.091)& 1.279& 1.375& 1.540\\
		&~0.362~(0.169) &~0.367~(0.145)&~0.373~(0.132)&~0.367~(0.131)& 1.166& 1.283& 1.289\\
		&~0.369~(0.169) &~0.364~(0.152)&~0.366~(0.137)&~0.370~(0.132)& 1.107& 1.230& 1.278\\
		&~0.508~(0.171) &~0.504~(0.151)&~0.508~(0.137)&~0.505~(0.133)& 1.131& 1.251& 1.282\\
		&~0.499~(0.164) &~0.499~(0.144)&~0.499~(0.137)&~0.499~(0.129)& 1.138& 1.200& 1.267\\
		\hline
		0.7 & 6.474~(0.193) & 6.483~(0.152)& 6.485~(0.139)& 6.485~(0.122)& 1.271& 1.387& 1.582\\
		& 0.520~(0.201) & 0.518~(0.166)& 0.518~(0.160)& 0.517~(0.153)& 1.207& 1.250& 1.311\\
		& 0.511~(0.188) & 0.513~(0.155)& 0.515~(0.153)& 0.510~(0.146)& 1.209& 1.229& 1.284\\
		& 0.515~(0.187) & 0.512~(0.160)& 0.510~(0.148)& 0.507~(0.144)& 1.169& 1.265& 1.301\\
		& 0.516~(0.190) & 0.512~(0.162)& 0.513~(0.152)& 0.510~(0.140)& 1.173& 1.250& 1.354\\
		\hline
		0.9 & 9.069~(0.294) & 9.099~(0.270)& 9.092~(0.267)& 9.096~(0.244)& 1.087& 1.100& 1.207\\
		& 0.646~(0.250) & 0.639~(0.245)& 0.645~(0.236)& 0.648~(0.219)& 1.021& 1.059& 1.140\\
		& 0.649~(0.270) & 0.658~(0.245)& 0.657~(0.231)& 0.654~(0.224)& 1.103& 1.172& 1.208\\
		& 0.512~(0.249) & 0.509~(0.228)& 0.502~(0.221)& 0.508~(0.210)& 1.092& 1.128& 1.188\\
		& 0.513~(0.253) & 0.503~(0.235)& 0.511~(0.223)& 0.510~(0.218)& 1.078& 1.135& 1.160\\
		\hline
	\end{tabular}
\end{table}

\begin{table}[htp]
	\footnotesize
	\caption{The AREs and means and standard deviations (in parentheses) of SL and SSL with different $\tau$s and Ns under $\bm{\varepsilon}\sim U(0,1)$.  }
	\centering
	\begin{tabular}{@{}c|c|ccc|cccc@{}}
		\hline
		&&&SSL&&&ARE\\
		\cline{3-8}
		$\tau$&SL&N=500&N=1000&N=2000&N=500&N=1000&N=2000\\
		\hline
		0.1 &2.640~(0.074) &2.626~(0.067)&2.622~(0.067)&2.616~(0.066)& 1.112& 1.116& 1.122\\
		&0.695~(0.147) &0.696~(0.134)&0.692~(0.130)&0.693~(0.129)& 1.100& 1.132& 1.140\\
		&0.690~(0.158) &0.694~(0.145)&0.691~(0.141)&0.696~(0.133)& 1.088& 1.125& 1.194\\
		&0.504~(0.157) &0.505~(0.146)&0.503~(0.145)&0.506~(0.143)& 1.075& 1.086& 1.099\\
		&0.507~(0.156) &0.504~(0.143)&0.506~(0.138)&0.504~(0.137)& 1.086& 1.129& 1.136\\
		\hline
		0.3 & 4.046~(0.098) & 4.042~(0.074)& 4.036~(0.067)& 4.038~(0.055)& 1.339& 1.460& 1.784\\
		& 0.720~(0.134) & 0.724~(0.115)& 0.720~(0.110)& 0.725~(0.106)& 1.165& 1.225& 1.264\\
		& 0.719~(0.142) & 0.716~(0.120)& 0.717~(0.110)& 0.719~(0.099)& 1.180& 1.283& 1.426\\
		& 0.494~(0.136) & 0.494~(0.124)& 0.494~(0.113)& 0.494~(0.107)& 1.096& 1.201& 1.268\\
		& 0.515~(0.143) & 0.506~(0.127)& 0.512~(0.115)& 0.512~(0.108)& 1.132& 1.241& 1.330\\
		\hline
		0.5 &~5.409~(0.129) &~5.419~(0.095)&~5.430~(0.078)&~5.436~(0.063)& 1.363& 1.666& 2.057\\
		&~0.730~(0.144) &~0.731~(0.125)&~0.735~(0.117)&~0.731~(0.112)& 1.154& 1.232& 1.288\\
		&~0.737~(0.152) &~0.735~(0.127)&~0.737~(0.121)&~0.735~(0.107)& 1.199& 1.256& 1.425\\
		&~0.502~(0.147) &~0.505~(0.129)&~0.502~(0.115)&~0.501~(0.110)& 1.136& 1.282& 1.334\\
		&~0.495~(0.156) &~0.494~(0.130)&~0.500~(0.118)&~0.498~(0.105)& 1.202& 1.317& 1.490\\
		\hline
		0.7 & 6.759~(0.186) & 6.784~(0.134)& 6.793~(0.120)& 6.806~(0.099)& 1.387& 1.554& 1.874\\
		& 0.749~(0.172) & 0.747~(0.142)& 0.751~(0.139)& 0.747~(0.128)& 1.213& 1.236& 1.339\\
		& 0.764~(0.167) & 0.759~(0.142)& 0.763~(0.132)& 0.758~(0.126)& 1.179& 1.270& 1.329\\
		& 0.497~(0.165) & 0.504~(0.146)& 0.504~(0.138)& 0.500~(0.127)& 1.128& 1.196& 1.297\\
		& 0.495~(0.167) & 0.495~(0.138)& 0.500~(0.127)& 0.492~(0.127)& 1.211& 1.320& 1.316\\
		\hline
		0.9 & 9.148~(0.284) & 9.183~(0.254)& 9.185~(0.238)& 9.181~(0.236)& 1.120& 1.194& 1.204\\
		& 0.765~(0.236) & 0.759~(0.222)& 0.768~(0.209)& 0.759~(0.200)& 1.066& 1.134& 1.183\\
		& 0.772~(0.237) & 0.773~(0.211)& 0.766~(0.209)& 0.775~(0.201)& 1.121& 1.134& 1.177\\
		& 0.497~(0.254) & 0.498~(0.230)& 0.506~(0.223)& 0.494~(0.215)& 1.105& 1.138& 1.181\\
		& 0.512~(0.222) & 0.502~(0.203)& 0.502~(0.196)& 0.496~(0.184)& 1.093& 1.135& 1.209\\
		\hline
	\end{tabular}
\end{table}

\subsection{Real data applications: the homeless data in Los Angeles County}
To illustrate the proposed methods, we used a data set, which are the total number of people estimated by the Los Angeles Homeless Services Administration (LAHSA) to be on the streets, shelters, or ``almost homeless'' in Los Angeles County from 2004 to 2005.
Because the entire County of Los Angeles includes 2,054 census tracts, it is expensive to survey the entire county, so stratified spatial sampling of census tracts is used.
It calls for two steps. In the first step, 244 tracts believed to have large numbers of homeless people were visited, known as ``hot tracts''. The second step was to visit a stratified random sample of tracts from the population of non-hot tracts. Stratified random sampling 265 tracts, the remaining 1545 tracts were not visited. Therefore, the dataset with total 1810 observations (labeled data 265 and unlabeled data 1545) is considered.
\par
We use the linear extremile regression model (1.5) to analyze the change in the homeless street count over the effect of four predictors,which are
PctVacant (\% of unoccupied housing units), PctOwnerOcc (\% of owner-occupied housing units), PctMinority (\% of population that is non-Caucasian) and
MedianHouseholdIncome (Median household income).
They are the important variable in \cite{r52}. The estimators of regression coefficients and theirs estimated standard deviations (estimated by the methods in section 4)
are present in Table 6. The results show that the proposed semi-supervised estimators
generally have significant improvement over the supervised estimators according to the smaller estimated standard deviations.

\begin{table}[htp]
	\footnotesize
	\caption{The estimators and theirs estimated standard deviations (Esd) of SL and SSL with different $\tau$s for the homeless data in Los Angeles County. ARE=Esd of SL/Esd of SSL.   }
	\centering
	\begin{tabular}{@{}c|c|cccccccccc@{}}
		\hline
		$\tau$&Method&Intercept&PctVacant&PctOwnerOcc&PctMinority&MedianHouseholdIncome\\
		\hline
		0.1 &Estimator of SL &6.908 &1.027&0.895&0.824&0.713\\
		&Estimator of SSL&7.213 &0.962&0.886&0.745&1.015\\
		&Esd of SL        &0.430 &0.394&0.430&0.412&0.376\\
		&Esd of SSL       &0.457 &0.368&0.416&0.375&0.350\\
		&ARE             &0.941 &1.073&1.033&1.101&1.077\\
		\hline
		0.3 &Estimator of SL &9.286 &0.757&0.811&0.515&0.833\\
		&Estimator of SSL&9.688 &0.794&0.914&0.472&1.207\\
		&Esd of SL        &0.558 &0.508&0.533&0.539&0.533\\
		&Esd of SSL       &0.572 &0.476&0.511&0.496&0.488\\
		&ARE             &0.975 &1.067&1.044&1.087&1.093\\
		\hline
		0.5 &Estimator of SL &11.352 &0.710&0.813&0.326&0.887\\
		&Estimator of SSL&11.815 &0.780&1.055&0.239&1.151\\
		&Esd of SL        &~0.608 &0.558&0.567&0.559&0.609\\
		&Esd of SSL       &~0.626 &0.522&0.543&0.520&0.539\\
		&ARE             &~0.971 &1.070&1.045&1.075&1.128\\
		\hline
		0.7 &Estimator of SL &13.407 &0.659&0.812&0.136&0.943\\
		&Estimator of SSL&13.933 &0.763&1.193&0.008&1.101\\
		&Esd of SL        &~0.674 &0.620&0.609&0.587&0.701\\
		&Esd of SSL       &~0.694 &0.579&0.582&0.552&0.606\\
		&ARE             &~0.972 &1.072&1.046&1.063&1.157\\
		\hline
		0.9 &Estimator of SL &16.703 &0.790&0.867&-0.089&0.983\\
		&Estimator of SSL&17.298 &0.915&1.488&-0.443&0.825\\
		&Esd of SL        &~0.752 &0.680&0.664&~0.581&0.810\\
		&Esd of SSL       &~0.776 &0.624&0.636&~0.552&0.675\\
		&ARE             &~0.968 &1.090&1.044&~1.053&1.201\\
		\hline
	\end{tabular}
\end{table}

\section{Conclusion}
This article introduces a novel approach to linear extremile regression, which does not rely on the estimation of an unknown distribution function like those in \cite{r1,r41}. Instead, it achieves $\sqrt{n}$-consistent estimators of unknown regression parameters using the parameter quantile method. This is of significant theoretical importance and aligns with expectations in parametric regression analysis. Furthermore, the proposed estimation methods can readily accommodate various $\tau$-extremiles, making them well-suited for big data analysis. Specifically, we explore semi-supervised settings and demonstrate, through Theorems 2.2 and 3.2 and simulation studies, the effectiveness of estimates obtained from both labeled and unlabeled data.
\par
Based on the research findings of this article, it becomes convenient to extend the application of ``Extremile'' into more complex models, such as single-index models \citep{r66} and varying coefficient models \citep{r69}, as well as with complex data types including massive data  \citep{r68} and streaming data, among others  \citep{r67}.

\section*{Acknowledgments}
This research is supported by the Humanities and Social Science Youth Foundation, Ministry of Education of the People's Republic of China, (Series number: 22YJC910005) and the National Social Science Foundation of China (Series number: 21BTU040).

\appendix
\section{Proof of Theorems} 
{\bf Proof of Proposition 2.1.} Based on (2.2) and (2.3), we can obtain that
\begin{equation*}
	\begin{split}
		\xi_{\tau}(\bm{X})&=\bm{X}^{\top}\bm{\beta}_{\tau}=
		\bm{X}^{\top}\int_0^1\bm{\gamma}_{\bar{\tau}}\text{J}_{\tau}
		(\bar{\tau})d\bar{\tau}=\int_0^1\bm{X}^{\top}\bm{\gamma}_{\bar{\tau}}
		\text{J}_{\tau}(\bar{\tau})d\bar{\tau}\\
		&=\int_0^1q_{\bar{\tau}}(\bm{X})\text{J}_{\tau}(\bar{\tau})d\bar{\tau}
		=\int_{y\in \mathbb{R}}y\text{J}_{\tau}(F(y|\bm{X}))dF(y|\bm{X})\\
		&=E\{\bm{Y}\text{J}_{\tau}(F(\bm{Y}|\bm{X}))\}=E(\bm{Z}_{\bm{X},{\tau}}),
	\end{split}
\end{equation*}
where $\bm{Z}_{\bm{X},{\tau}}$ has cumulative distribution function $F_{\bm{Z}_{\bm{X},{\tau}}}=\text{K}_{\tau}(F(\cdot|\bm{X}))$.
When $\tau=0.5^{1/r}$ and $r\in \mathbb{N}\backslash \{0\}$, for any $z\in \mathbb{R}$, we have
\begin{equation*}
	\begin{split}
		F_{\bm{Z}_{\bm{X},{\tau}}}(z)=\text{K}_{\tau}(F(z|\bm{X}))=\{F(z|\bm{X})\}^r=P\left(\max(Y_{\bm{X}}^1,\ldots,Y_{\bm{X}}^{r})\leqslant z\right),
	\end{split}
\end{equation*}
so that $\xi_{\tau}(\bm{X})
=E(\bm{Z}_{\bm{X},{\tau}})=E\{\max(Y_{\bm{X}}^1,\ldots,Y_{\bm{X}}^{r})\}$ according to
$\bm{Z}_{\bm{X},{\tau}}=\max(Y_{\bm{X}}^1,\ldots,Y_{\bm{X}}^{r})$. Similarly, we can prove that $\xi_{\tau}(\bm{X})
=E\{\min(Y_{\bm{X}}^1,\ldots,Y_{\bm{X}}^{r})\}$ under $\tau=1-0.5^{1/s}$ and $s\in \mathbb{N}\backslash \{0\}$. Because,
for any $z\in \mathbb{R}$,
 \begin{equation*}
 	\begin{split}
 		1-F_{\bm{Z}_{\bm{X},{\tau}}}(z)=1-\text{K}_{\tau}(F(z|\bm{X}))=\{1-F(z|\bm{X})\}^s
 		=P\left(\min(Y_{\bm{X}}^1,\ldots,Y_{\bm{X}}^{r})> z\right).
 	\end{split}
 \end{equation*}
 \\
\par
{\bf Proof of Theorems 2.1 and 2.2.} The results of Theorems 2.1 and 2.2 can be obtained directly from the following proofs of theorems 3.1 and 3.2 under $N=0$ and $\bm{\alpha}^*=\bm{\alpha}_0$.
\\
\par
{\bf Proof of Theorem 3.1}.
{\bf To establish consistency.}
Denote
\begin{equation*}
	\begin{split}
		\tilde{L}(\bm{\alpha})=\sum_{i=1}^{n}L(Y_i,\bm{X}_i,\bm{\alpha})
		+\sum_{i=n+1}^{n+N}\bm{Z}_i^{\top}\hat{\varphi}(\bm{\alpha}).
	\end{split}
\end{equation*}
Note that $\tilde{L}(\bm{\alpha})$ is the loss function in equation (3.3). We first show that
$\tilde{L}(\bm{\alpha})$ is invariant under affine transformation on $\bm{Z}=(1,\tilde{\bm{Z}}^{\top})^{\top}$, where $\tilde{\bm{Z}}$ is the remainder of $\bm{Z}$ after removing the first element. Assume that $\tilde{\bm{Z}}_i=\bm{M}\tilde{\bm{Q}}_i+\bm{b}$ for any fixed nonsingular $(d-1)\times (d-1)$ matrix $\bm{M}$ and $d-1$ vector $\bm{b}$, where $\tilde{\bm{Z}}_i$ is the $i$-th observation of $\tilde{\bm{Z}}$. Then, we can rewrite $\bm{Z}_i$ as
$$
\bm{Z}_i=\left(
\begin{array}{ll}
	1&{\bf 0}^{\top}_{q-1}\\
	\bm{b}&\bm{M}
\end{array}
\right)
\bm{Q}_i,
$$
where $\bm{Q}_i=(1,\tilde{\bm{Q}}_i^{\top})^{\top}$. Then, for any $\bm{\alpha}$, we can obtain
\begin{equation*}
	\begin{split}
		\tilde{L}(\bm{\alpha})=&
		\sum_{i=1}^{n}L(Y_i,\bm{X}_i,\bm{\alpha})
		+\sum_{i=n+1}^{n+N}\bm{Z}_i^{\top}
		\left(\frac{1}{n}\sum_{i=1}^{n}\bm{Z}_i\bm{Z}_i^{\top}\right)^{-1}
		\frac{1}{n}\sum_{i=1}^{n}\bm{Z}_iL(Y_i,\bm{X}_i,\bm{\alpha})\\
		=&\sum_{i=1}^{n}L(Y_i,\bm{X}_i,\bm{\alpha})
		+\sum_{i=n+1}^{n+N}\bm{Q}_i^{\top}
		\left(\frac{1}{n}\sum_{i=1}^{n}\bm{Q}_i\bm{Q}_i^{\top}\right)^{-1}
		\frac{1}{n}\sum_{i=1}^{n}\bm{Q}_iL(Y_i,\bm{X}_i,\bm{\alpha}).
	\end{split}
\end{equation*}
Therefore, we consider $E(\bm{Z}\bm{Z}^{\top})=\bm{I}_d$ and $E(\bm{Z})=(1,\bm{0}_{d-1}^{\top})^{\top}$ in the following proofs according to the above affine transformation invariant property.
Let
\begin{equation}
	\begin{split}
		\bar{L}(\bm{\alpha})=&\sum_{i=1}^{n}L(Y_i,\bm{X}_i,\bm{\alpha})
		+NE(\bm{Z}^{\top})\{E(\bm{Z}\bm{Z}^{\top})\}^{-1}
		\frac{1}{n}\sum_{i=1}^{n}\bm{Z}_iL(Y_i,\bm{X}_i,\bm{\alpha})\\
		=&\frac{n+N}{n}\sum_{i=1}^{n}L(Y_i,\bm{X}_i,\bm{\alpha}).
	\end{split}
\end{equation}
\par
Next, we proof that $\sup_{\bm{\alpha}\in\Theta}|\tilde{L}(\bm{\alpha})-\bar{L}(\bm{\alpha})|=o_p(1)$. Then, by Lemma 1 in \cite{r48} and conditions {\bf C2} and {\bf C5}, for large enough constants $c_1$ and $c_2$, we have
\begin{equation}
	\begin{split}
		&|\tilde{L}(\bm{\alpha})-\bar{L}(\bm{\alpha})|\\
		=&
		\left|\left[\sum_{i=n+1}^{n+N}\bm{Z}_i^{\top}
		\left(\frac{1}{n}\sum_{i=1}^{n}\bm{Z}_i\bm{Z}_i^{\top}\right)^{-1}
		-NE(\bm{Z}^{\top})\{E(\bm{Z}\bm{Z}^{\top})\}^{-1}\right]
		\frac{1}{n}\sum_{i=1}^{n}\bm{Z}_iL(Y_i,\bm{X}_i,\bm{\alpha})\right|\\
		\leq&\left\|\sum_{i=n+1}^{n+N}\bm{Z}_i^{\top}
		\left(\frac{1}{n}\sum_{i=1}^{n}\bm{Z}_i\bm{Z}_i^{\top}\right)^{-1}
		-NE(\bm{Z}^{\top})\{E(\bm{Z}\bm{Z}^{\top})\}^{-1}\right\|_2
		\left\|\frac{1}{n}\sum_{i=1}^{n}\bm{Z}_iL(Y_i,\bm{X}_i,\bm{\alpha})\right\|_2\\
		\leq&c_1\left\|\left\{\sum_{i=n+1}^{n+N}\bm{Z}_i^{\top}-NE(\bm{Z}^{\top})\right\}
		\left(\sum_{i=1}^{n}\bm{Z}_i\bm{Z}_i^{\top}\right)^{-1}\right\|_2\\
		&+c_1N\left\|E(\bm{Z}^{\top})
		\left[\left(\frac{1}{n}\sum_{i=1}^{n}\bm{Z}_i\bm{Z}_i^{\top}\right)^{-1}
		-E(\bm{Z}^{\top})\{E(\bm{Z}\bm{Z}^{\top})\}^{-1}\right]\right\|_2\\
		\leq&c_2(N^{1/2}+Nn^{-1/2}),
	\end{split}
\end{equation}
where $\|\cdot\|_2$ is $L_2$ norm. Then, by equations (A.1) and (A.2), and $n\rightarrow\infty$, we have
\begin{equation}
	\begin{split}
		&\sup_{\bm{\alpha}\in\Theta}|\tilde{L}(\bm{\alpha})/(n+N)
		-E\{L(\bm{Y},\bm{X},\bm{\alpha})\}|\\
		&\leq \sup_{\bm{\alpha}\in\Theta}|\tilde{L}(\bm{\alpha})-\bar{L}(\bm{\alpha})|/(n+N)
		+\sup_{\bm{\alpha}\in\Theta}|\bar{L}(\bm{\alpha})/(n+N)-E\{L(\bm{Y},\bm{X},\bm{\alpha})\}|
		=o_p(1).
	\end{split}
\end{equation}
(A.3) implies that $\|\text{Vec}(\tilde{\bm{\alpha}}-\bm{\alpha}^*)\|_2=o_p(1)$ according to $\bm{\alpha}^*$ is the unique minimizer of $E\{L(\bm{Y},\bm{X},\bm{\alpha})\}$.  Thus, the consistency is proved.
\\
\par
{\bf To show asymptotic normality.} By equation (3.3) and Taylor's expansion of $\nabla_{\text{Vec}(\bm{\alpha})}\tilde{L}(\bm{\alpha})|_{\bm{\alpha}=\tilde{\bm{\alpha}}}$ at $\bm{\alpha}^*$ as
\begin{equation}
	\begin{split}
		\bm{0}=\nabla_{\text{Vec}(\bm{\alpha})}\tilde{L}(\bm{\alpha})|_{\bm{\alpha}=\tilde{\bm{\alpha}}}
		=\nabla_{\text{Vec}(\bm{\alpha})}\tilde{L}(\bm{\alpha})|_{\bm{\alpha}=\bm{\alpha}^*}+
		\nabla^2_{\text{Vec}(\bm{\alpha})}
		\tilde{L}(\bm{\alpha})|_{\bm{\alpha}=\bar{\bm{\alpha}}}\text{Vec}(\tilde{\bm{\alpha}}
		-\bm{\alpha}^*),
	\end{split}
\end{equation}
where $\bar{\bm{\alpha}}$ is between $\tilde{\bm{\alpha}}$ and $\bm{\alpha}^*$.
We first consider $\nabla_{\text{Vec}(\bm{\alpha})}\tilde{L}(\bm{\alpha})|_{\bm{\alpha}=\bm{\alpha}^*}$.
Denote $\tilde{\bm{Z}}_N=\sum_{i=n+1}^{n+N}\bm{Z}_i/N$, $\tilde{\bm{Z}}_n=\sum_{i=1}^{n}\bm{Z}_i/n$, $\hat{{\bm \Sigma}}_{\bm{Z}}=\sum_{i=1}^{n}\bm{Z}_i\bm{Z}_i^{\top}/n$ and
$\bm{U}_i=S_i(\bm{\alpha}^*)-\bm{A}^{\top}\bm{Z}_i$ with $S_i(\bm{\alpha}^*)=\nabla_{\text{Vec}(\bm{\alpha})}L(Y_i,\bm{X}_i,\bm{\alpha})|_{\bm{\alpha}=\bm{\alpha}^*}$. Thus, we can obtain
\begin{equation}
	\begin{split}
		\nabla_{\text{Vec}(\bm{\alpha})}\tilde{L}(\bm{\alpha})|_{\bm{\alpha}=\bm{\alpha}^*}
		=&nS(\bm{\alpha}^*)+\frac{N}{n}\sum_{i=1}^{n}S_i(\bm{\alpha}^*)\bm{Z}_i^{\top}
		\hat{{\bm \Sigma}}_{\bm{Z}}^{-1}\tilde{\bm{Z}}_N\\
		=&nS(\bm{\alpha}^*)+\frac{N}{n}
		\sum_{i=1}^{n}(\bm{A}^{\top}\bm{Z}_i+\bm{U}_i)\bm{Z}_i^{\top}\hat{{\bm \Sigma}}_{\bm{Z}}^{-1}\tilde{\bm{Z}}_N\\
		=&nS(\bm{\alpha}^*)+N\bm{A}^{\top}\left\{\tilde{\bm{Z}}_N-E(\bm{Z})\right\}
		+N\bm{A}^{\top}E(\bm{Z})\\
		&+\frac{N}{n}\sum_{i=1}^{n}\bm{U}_i\bm{Z}_i^{\top}
		\hat{{\bm \Sigma}}_{\bm{Z}}^{-1}\tilde{\bm{Z}}_n+\frac{N}{n}\sum_{i=1}^{n}\bm{U}_i\bm{Z}_i^{\top}
		\hat{{\bm \Sigma}}_{\bm{Z}}^{-1}
		\left\{\tilde{\bm{Z}}_N-\tilde{\bm{Z}}_n\right\}\\
		=&\left\{nS(\bm{\alpha}^*)+\frac{N}{n}\sum_{i=1}^{n}\bm{U}_i\right\}
		+N\bm{A}^{\top}\left\{\tilde{\bm{Z}}_N-E(\bm{Z})\right\}\\
		&+o_p(n^{-1/2}(n+N))\\
		=&\left\{(n+N)S(\bm{\alpha}^*)-N\bm{A}^{\top}\tilde{\bm{Z}}_n\right\}
		+N\bm{A}^{\top}\left\{\tilde{\bm{Z}}_N-E(\bm{Z})\right\}\\
		&+o_p(n^{-1/2}(n+N)),
	\end{split}
\end{equation}
where
the forth equality holds because of $\bm{A}^{\top}E(\bm{Z})=\bm{0}$ by the definition of $\bm{\alpha}^*$, $
\hat{{\bm \Sigma}}_{\bm{Z}}^{-1}\tilde{\bm{Z}}_n=(1,\bm{0}_{d-1}^{\top})^{\top}$ by Lemma 2 in \cite{r47}, and $$\frac{N}{n}\sum_{i=1}^{n}\bm{U}_i\bm{Z}_i^{\top}
\hat{{\bm \Sigma}}_{\bm{Z}}^{-1}
\left\{\tilde{\bm{Z}}_N-\tilde{\bm{Z}}_n\right\}=o_p(n^{-1/2}(n+N))$$ by proof similar to Theorem 1 in \cite{r47}.
\par
Finally, we consider $\nabla^2_{\text{Vec}(\bm{\alpha})}\tilde{L}(\bm{\alpha})|_{\bm{\alpha}=\bar{\bm{\alpha}}}$ as
\begin{equation}
	\begin{split}
		\nabla^2_{\text{Vec}(\bm{\alpha})}\tilde{L}(\bm{\alpha})|_{\bm{\alpha}=\bar{\bm{\alpha}}}
		=&
		\nabla^2_{\text{Vec}(\bm{\alpha})}\bar{L}(\bm{\alpha})|_{\bm{\alpha}=\bar{\bm{\alpha}}}
		+\left\{\nabla^2_{\text{Vec}(\bm{\alpha})}\tilde{L}(\bm{\alpha})|_{\bm{\alpha}=\bar{\bm{\alpha}}}
		-\nabla^2_{\text{Vec}(\bm{\alpha})}\bar{L}(\bm{\alpha})|_{\bm{\alpha}=\bar{\bm{\alpha}}} \right\}
		\\
		=&(n+N){\bf H}+\left\{\nabla^2_{\text{Vec}(\bm{\alpha})}\bar{L}(\bm{\alpha})|_{\bm{\alpha}
			=\bar{\bm{\alpha}}}-(n+N){\bf H}\right\}
		+O_p(N^{1/2}+Nn^{-1/2})\\
		=&(n+N)\{{\bf H}+o_p(1)\},
	\end{split}
\end{equation}
where the second equation is similar to (A.2) by conditions {\bf C2} and {\bf C4}, and the last equation is according to (A.1). Then, from (A.4)-(A.6), we have
\begin{equation}
	\begin{split}
		\text{Vec}(\tilde{\bm{\alpha}}-\bm{\alpha}^*)
		=-{\bf H}^{-1}\left[\left\{S(\bm{\alpha}^*)-\frac{N}{n+N}\bm{A}^{\top}\tilde{\bm{Z}}_n
		\right\}
		+\frac{N}{n+N}\bm{A}^{\top}\left\{\tilde{\bm{Z}}_N-E(\bm{Z})\right\}
		\right]+o_p(n^{-1/2}).
	\end{split}
\end{equation}
Therefore, we can prove the theorem.
\\
\par
{\bf Proof of Theorem 3.2}. From the $\tilde{\beta}_{\tau}=\tilde{\alpha}\int_0^1\bm{b}(\bar{\tau})\text{J}_{\tau}(\bar{\tau})d\bar{\tau}$ and (A.7), the theorem can be directly proven.

\section{The asymptotic property of $\bar{\bm{\beta}}_{\tau}$ in the equation (1.7)}
In this section, we first assume that $\text{F}(\cdot|\bm{X})$ is known. Then, we can estimate $\bm{\beta}_{\tau}$ in the linear extremile regression (1.5) as
\begin{equation}
	\begin{split}
		\bar{\bm{\beta}}^*_{\tau}=&\arg\min_{\bm{\beta}}\sum_{i=1}^{n}
		\text{J}_{\tau}(\text{F}(Y_i|\bm{X}_i))\cdot(Y_i-\bm{X}_i^{\top}\bm{\beta})^2\\
		=&(\bm{X}^{\top}\bm{W}\bm{X})^{-1}\bm{X}^{\top}\bm{W}\bm{Y},
	\end{split}
\end{equation}
where $\bm{W}=\text{diag}\{\text{J}_{\tau}(\text{F}(Y_1|\bm{X}_1)),\ldots,
\text{J}_{\tau}(\text{F}(Y_n|\bm{X}_n))\}$.

\begin{thm}
	Suppose that  $\bm{\Sigma}_{J\bm{X}}=E\{\text{J}_{\tau}(\text{F}(\bm{Y}|\bm{X}))\bm{X}^{\top}\bm{X}\}$ is nonsingular and $n\rightarrow\infty$, we have
	\begin{equation*}
		\begin{split}
			\sqrt{n}(\bar{\bm{\beta}}^*_{\tau}-\bm{\beta}_{\tau})
			\xrightarrow{L}\textrm{N}\left(\bm{0},\bm{\Sigma}_{J\bm{X}}^{-1}{\bf G}\bm{\Sigma}_{J\bm{X}}^{-1}\right),
		\end{split}
	\end{equation*}
	where ${\bf G}=E\{\text{J}_{\tau}(\text{F}(\bm{Y}|\bm{X}))^2(\bm{Y}-\bm{X}^{\top}\bm{\beta}_{\tau})^2\bm{X}^{\top}\bm{X}\}$.
\end{thm}
\par
{\bf Proof of Theorem B.1}.
From the equation (B.1), we have
\begin{equation}
	\begin{split}
		\sqrt{n}(\bar{\bm{\beta}}^*_{\tau}-\bm{\beta}_{\tau})
		=&(\bm{X}^{\top}\bm{W}\bm{X})^{-1}\bm{X}^{\top}\bm{W}(\bm{Y}-\bm{X}^{\top}\bm{\beta}_{\tau})\\
		=&\{\bm{\Sigma}_{J\bm{X}}+o_p(1)\}^{-1}\frac{1}{\sqrt{n}}\sum_{i=1}^{n}\text{J}_{\tau}(\text{F}(Y_i|\bm{X}_i))\bm{X}_i^{\top}(Y_i-\bm{X}_i^{\top}\bm{\beta}_{\tau}).
	\end{split}
\end{equation}
\par
From the definition of $\bm{\beta}_{\tau}$ in (1.6), we have
\begin{equation}
	\begin{split}
		\text{E}\left\{\text{J}_{\tau}(\text{F}(\bm{Y|\bm{X}}))\bm{X}^{\top}(\bm{Y}-\bm{X}^{\top}\bm{\beta}_{\tau})\right\}={\bf 0}.
	\end{split}
\end{equation}
Therefore, Theorem B.1 can be proved by the Central limit theorem based on (B.2) and (B.3).
\newpage
\par
We now consider the asymptotic property of $\bar{\bm{\beta}}_{\tau}$ in the equation (1.7).
\begin{thm}
	Suppose that  $\bm{\Sigma}_{J\bm{X}}=E\{\text{J}_{\tau}(\text{F}(\bm{Y}|\bm{X}))\bm{X}^{\top}\bm{X}\}$ is nonsingular and $\|E\{(\bm{Y}-\bm{X}^{\top}\bm{\beta}_{\tau})^2\bm{X}^{\top}\bm{X}\}\|<\infty$. If
	$\text{F}(y|\bm{X})$ and the marginal density of $\bm{X}$ have continuous second-order partial derivatives with respect to $\bm{X}$,
	the kernel $K(\cdot)$ is a symmetric density function with finite support		
	and $h=O(n^{-1/5})$. Then as $n\rightarrow\infty$, we have
	\begin{equation*}
		\begin{split}
			\bar{\bm{\beta}}_{\tau}-\bm{\beta}_{\tau}=O_p(n^{-2/5}),
		\end{split}
	\end{equation*}
	where $\|\cdot\|$ is the spectral norm.
\end{thm}
\par
{\bf Proof of Theorem B.2}. 	
Based on (1.7), we can obtain
\begin{equation}
	\begin{split}
		\bar{\bm{\beta}}_{\tau}-\bm{\beta}_{\tau}
		=&(\bm{X}^{\top}\hat{\bm{W}}\bm{X})^{-1}\bm{X}^{\top}\hat{\bm{W}}\bm{Y}\\
		=&\{\bm{\Sigma}_{J\bm{X}}+o_p(1)\}^{-1}\frac{1}{n}\sum_{i=1}^{n}\text{J}_{\tau}(\hat{\text{F}}(Y_i|\bm{X}_i))\bm{X}_i^{\top}(Y_i-\bm{X}_i^{\top}\bm{\beta}_{\tau})\\
		=&(\bar{\bm{\beta}}^*_{\tau}-\bm{\beta}_{\tau})+\{\bm{\Sigma}_{J\bm{X}}+o_p(1)\}^{-1}\frac{1}{n}\sum_{i=1}^{n}\left\{\text{J}_{\tau}(\hat{\text{F}}(Y_i|\bm{X}_i))-\text{J}_{\tau}(\text{F}(Y_i|\bm{X}_i))
		\right\}\bm{X}_i^{\top}(Y_i-\bm{X}_i^{\top}\bm{\beta}_{\tau}).
	\end{split}
\end{equation}
Based on (1.8) and the conditions in Theorem B.2, we can obtain
\begin{equation}
	\begin{split}
		\hat{\text{F}}(y|\bm{x})-\text{F}(y|\bm{x})=O_p(n^{-2/5}).
	\end{split}
\end{equation}
Moreover, by the Theorem B.1, we have
\begin{equation}
	\begin{split}
		\bar{\bm{\beta}}^*_{\tau}-\bm{\beta}_{\tau}=O_p(n^{-1/2}).
	\end{split}
\end{equation}
Therefore, by (B.4)-(B.6) and condition $\|E\{(\bm{Y}-\bm{X}^{\top}\bm{\beta}_{\tau})^2\bm{X}^{\top}\bm{X}\}\|<\infty$, we can prove this theorem.
\\
\\
\\
\bibliographystyle{imsart-nameyear}
\bibliography{ref}

\begin{thebibliography}{33}

\bibitem[\protect\citeauthoryear{Azriel et~al.}{2022}]{r63}
\begin{barticle}[author]
\bauthor{\bsnm{Azriel},~\bfnm{D}\binits{D.}},
  \bauthor{\bsnm{Brown},~\bfnm{L.~D}\binits{L.~D.}},
  \bauthor{\bsnm{Sklar},~\bfnm{M}\binits{M.}},
  \bauthor{\bsnm{Berk},~\bfnm{R}\binits{R.}},
  \bauthor{\bsnm{Buja},~\bfnm{A}\binits{A.}}, \bauthor{} \AND
  \bauthor{\bsnm{Zhao},~\bfnm{L}\binits{L.}}
(\byear{2022}).
\btitle{Semi-supervised linear regression}.
\bjournal{Journal of the American Statistical Association}
\bvolume{117}
\bpages{2238-2251}.
\bdoi{10.1080/01621459.2021.1915320}
\end{barticle}
\endbibitem

\bibitem[\protect\citeauthoryear{Cai and Guo}{2020}]{r62}
\begin{barticle}[author]
\bauthor{\bsnm{Cai},~\bfnm{T.}\binits{T.}} \AND
  \bauthor{\bsnm{Guo},~\bfnm{Zijian}\binits{Z.}}
(\byear{2020}).
\btitle{Semisupervised inference for explained variance in high dimensional
  linear regression and its applications}.
\bjournal{Journal of the Royal Statistical Society: Series B (Statistical
  Methodology)}
\bvolume{82}
\bpages{391-419}.
\bdoi{10.1111/rssb.12357}
\end{barticle}
\endbibitem

\bibitem[\protect\citeauthoryear{Cannings}{2021}]{r65}
\begin{barticle}[author]
\bauthor{\bsnm{Cannings},~\bfnm{Timothy}\binits{T.}}
(\byear{2021}).
\btitle{Random projections: Data perturbation for classification problems}.
\bjournal{WIREs Computational Statistics}
\bvolume{13}
\bpages{e1499}.
\bdoi{10.1002/wics.1499}
\end{barticle}
\endbibitem

\bibitem[\protect\citeauthoryear{Chakrabortty and Cai}{2018}]{r64}
\begin{barticle}[author]
\bauthor{\bsnm{Chakrabortty},~\bfnm{Abhishek}\binits{A.}} \AND
  \bauthor{\bsnm{Cai},~\bfnm{Tianxi}\binits{T.}}
(\byear{2018}).
\btitle{Efficient and adaptive linear regression in semi-supervised settings}.
\bjournal{Annals of Statistics}
\bvolume{46}
\bpages{1541-1572}.
\bdoi{10.1214/17-AOS1594}
\end{barticle}
\endbibitem

\bibitem[\protect\citeauthoryear{Chapelle, Sch$\"{o}$lkopf and
  Zien}{2010}]{r54}
\begin{barticle}[author]
\bauthor{\bsnm{Chapelle},~\bfnm{Olivier}\binits{O.}},
  \bauthor{\bsnm{Sch$\"{o}$lkopf},~\bfnm{Bernhard}\binits{B.}} \AND
  \bauthor{\bsnm{Zien},~\bfnm{Alexander}\binits{A.}}
(\byear{2010}).
\btitle{Semi-supervised learning}.
\bjournal{the MIT Press}.
\end{barticle}
\endbibitem

\bibitem[\protect\citeauthoryear{Chen, Ma and Sun}{2023}]{r50}
\begin{barticle}[author]
\bauthor{\bsnm{Chen},~\bfnm{Yu}\binits{Y.}},
  \bauthor{\bsnm{Ma},~\bfnm{Mengyuan}\binits{M.}} \AND
  \bauthor{\bsnm{Sun},~\bfnm{Hongfang}\binits{H.}}
(\byear{2023}).
\btitle{Statistical inference for extreme extremile in heavy-tailed
  heteroscedastic regression model}.
\bjournal{Insurance: Mathematics and Economics}
\bvolume{111}
\bpages{142-162}.
\bdoi{10.1016/j.insmatheco.2023.04.001}
\end{barticle}
\endbibitem

\bibitem[\protect\citeauthoryear{Cheplygina, de~Bruijne and Pluim}{2019}]{r56}
\begin{barticle}[author]
\bauthor{\bsnm{Cheplygina},~\bfnm{Veronika}\binits{V.}},
  \bauthor{\bparticle{de} \bsnm{Bruijne},~\bfnm{Marleen}\binits{M.}} \AND
  \bauthor{\bsnm{Pluim},~\bfnm{Josien}\binits{J.}}
(\byear{2019}).
\btitle{Not-so-supervised: A survey of semi-supervised, multi-instance, and
  transfer learning in medical image analysis}.
\bjournal{Medical Image Analysis}
\bvolume{54}
\bpages{280-296}.
\bdoi{10.1016/j.media.2019.03.009}
\end{barticle}
\endbibitem

\bibitem[\protect\citeauthoryear{Daouia, Gijbels and Stupfler}{2019}]{r1}
\begin{barticle}[author]
\bauthor{\bsnm{Daouia},~\bfnm{Abdelaati}\binits{A.}},
  \bauthor{\bsnm{Gijbels},~\bfnm{Irene}\binits{I.}} \AND
  \bauthor{\bsnm{Stupfler},~\bfnm{Gilles}\binits{G.}}
(\byear{2019}).
\btitle{Extremiles: A new perspective on asymmetric least squares}.
\bjournal{Journal of the American Statistical Association}
\bvolume{114}
\bpages{1366-1381}.
\bdoi{10.1080/01621459.2018.1498348}
\end{barticle}
\endbibitem

\bibitem[\protect\citeauthoryear{Daouia, Gijbels and Stupfler}{2022}]{r41}
\begin{barticle}[author]
\bauthor{\bsnm{Daouia},~\bfnm{Abdelaati}\binits{A.}},
  \bauthor{\bsnm{Gijbels},~\bfnm{Irene}\binits{I.}} \AND
  \bauthor{\bsnm{Stupfler},~\bfnm{Gilles}\binits{G.}}
(\byear{2022}).
\btitle{Extremile regression}.
\bjournal{Journal of the American Statistical Association}
\bvolume{117}
\bpages{1579-1586}.
\bdoi{10.1080/01621459.2021.1875837}
\end{barticle}
\endbibitem

\bibitem[\protect\citeauthoryear{Flutre et~al.}{2013}]{r61}
\begin{barticle}[author]
\bauthor{\bsnm{Flutre},~\bfnm{Timothée}\binits{T.}},
  \bauthor{\bsnm{Wen},~\bfnm{Xiaoquan}\binits{X.}},
  \bauthor{\bsnm{Pritchard},~\bfnm{Jonathan}\binits{J.}} \AND
  \bauthor{\bsnm{Stephens},~\bfnm{Matthew}\binits{M.}}
(\byear{2013}).
\btitle{A Statistical Framework for Joint eQTL Analysis in Multiple Tissues}.
\bjournal{PLoS genetics}
\bvolume{9}
\bpages{e1003486}.
\bdoi{10.1371/journal.pgen.1003486}
\end{barticle}
\endbibitem

\bibitem[\protect\citeauthoryear{Frumento and Bottai}{2016}]{r44}
\begin{barticle}[author]
\bauthor{\bsnm{Frumento},~\bfnm{Paolo}\binits{P.}} \AND
  \bauthor{\bsnm{Bottai},~\bfnm{Matteo}\binits{M.}}
(\byear{2016}).
\btitle{Parametric modeling of quantile regression coefficient functions}.
\bjournal{Biometrics}
\bvolume{72}
\bpages{74-84}.
\bdoi{10.1111/biom.12410}
\end{barticle}
\endbibitem

\bibitem[\protect\citeauthoryear{Frumento, Bottai and
  Fern\'{a}ndez-Val}{2021}]{r45}
\begin{barticle}[author]
\bauthor{\bsnm{Frumento},~\bfnm{Paolo}\binits{P.}},
  \bauthor{\bsnm{Bottai},~\bfnm{Matteo}\binits{M.}} \AND
  \bauthor{\bsnm{Fern\'{a}ndez-Val},~\bfnm{Iv\'{a}n}\binits{I.}}
(\byear{2021}).
\btitle{Parametric modeling of quantile regression coefficient functions with
  longitudinal data}.
\bjournal{Journal of the American Statistical Association}
\bvolume{116}
\bpages{783-797}.
\bdoi{10.1080/01621459.2021.1892702}
\end{barticle}
\endbibitem

\bibitem[\protect\citeauthoryear{Huang et~al.}{2014}]{huang2014semi}
\begin{barticle}[author]
\bauthor{\bsnm{Huang},~\bfnm{Gao}\binits{G.}},
  \bauthor{\bsnm{Song},~\bfnm{Shiji}\binits{S.}},
  \bauthor{\bsnm{Gupta},~\bfnm{Jatinder~ND}\binits{J.~N.}} \AND
  \bauthor{\bsnm{Wu},~\bfnm{Cheng}\binits{C.}}
(\byear{2014}).
\btitle{Semi-supervised and unsupervised extreme learning machines}.
\bjournal{IEEE transactions on cybernetics}
\bvolume{44}
\bpages{2405-2417}.
\end{barticle}
\endbibitem

\bibitem[\protect\citeauthoryear{Jiang, Hu and Yu}{2022}]{r66}
\begin{barticle}[author]
\bauthor{\bsnm{Jiang},~\bfnm{Rong}\binits{R.}},
  \bauthor{\bsnm{Hu},~\bfnm{Xueping}\binits{X.}} \AND
  \bauthor{\bsnm{Yu},~\bfnm{Keming}\binits{K.}}
(\byear{2022}).
\btitle{Single-index expectile models for estimating conditional value at risk
  and expected shortfall}.
\bjournal{Journal of Financial Econometrics}
\bvolume{20}
\bpages{345-366}.
\bdoi{10.1093/jjfinec/nbaa016}
\end{barticle}
\endbibitem

\bibitem[\protect\citeauthoryear{Koenker and Bassett}{1978}]{r5}
\begin{barticle}[author]
\bauthor{\bsnm{Koenker},~\bfnm{Roger}\binits{R.}} \AND
  \bauthor{\bsnm{Bassett},~\bfnm{Gilbert}\binits{G.}}
(\byear{1978}).
\btitle{Regression quantile}.
\bjournal{Econometrica}
\bvolume{46}
\bpages{33-50}.
\bdoi{10.2307/1913643}
\end{barticle}
\endbibitem

\bibitem[\protect\citeauthoryear{Kriegler and Berk}{2010}]{r52}
\begin{barticle}[author]
\bauthor{\bsnm{Kriegler},~\bfnm{Brian}\binits{B.}} \AND
  \bauthor{\bsnm{Berk},~\bfnm{Richard}\binits{R.}}
(\byear{2010}).
\btitle{Small area estimation of the homeless in Los Angeles: an application of
  cost-sensitive stochastic gradient boosting}.
\bjournal{Annals of Applied Statistics}
\bvolume{4}
\bpages{1234-1255}.
\bdoi{10.1214/10-AOAS328}
\end{barticle}
\endbibitem

\bibitem[\protect\citeauthoryear{LeBlanc, Moon and Kooperberg}{2006}]{r71}
\begin{barticle}[author]
\bauthor{\bsnm{LeBlanc},~\bfnm{Michael}\binits{M.}},
  \bauthor{\bsnm{Moon},~\bfnm{James}\binits{J.}} \AND
  \bauthor{\bsnm{Kooperberg},~\bfnm{Charles}\binits{C.}}
(\byear{2006}).
\btitle{Extreme regression}.
\bjournal{Biostatistics}
\bvolume{7}
\bpages{71-84}.
\bdoi{10.1093/biostatistics/kxi041}
\end{barticle}
\endbibitem

\bibitem[\protect\citeauthoryear{Lv, Guo and Wu}{2022}]{r49}
\begin{barticle}[author]
\bauthor{\bsnm{Lv},~\bfnm{Jiang}\binits{J.}},
  \bauthor{\bsnm{Guo},~\bfnm{Chaohui}\binits{C.}} \AND
  \bauthor{\bsnm{Wu},~\bfnm{Jibo}\binits{J.}}
(\byear{2022}).
\btitle{Jackknife partially linear model averaging for the conditional quantile
  prediction}.
\end{barticle}
\endbibitem

\bibitem[\protect\citeauthoryear{Ma, Lin and Gai}{2022}]{r67}
\begin{barticle}[author]
\bauthor{\bsnm{Ma},~\bfnm{Xiaoyu}\binits{X.}},
  \bauthor{\bsnm{Lin},~\bfnm{Lu}\binits{L.}} \AND
  \bauthor{\bsnm{Gai},~\bfnm{Yujie}\binits{Y.}}
(\byear{2022}).
\btitle{A general framework of online updating variable selection for
  generalized linear models with streaming datasets}.
\bjournal{Journal of Statistical Computation and Simulation}.
\bdoi{10.1080/00949655.2022.2107207}
\end{barticle}
\endbibitem

\bibitem[\protect\citeauthoryear{Michaelson, Loguercio and Beyer}{2009}]{r59}
\begin{barticle}[author]
\bauthor{\bsnm{Michaelson},~\bfnm{Jacob}\binits{J.}},
  \bauthor{\bsnm{Loguercio},~\bfnm{Salvatore}\binits{S.}} \AND
  \bauthor{\bsnm{Beyer},~\bfnm{Andreas}\binits{A.}}
(\byear{2009}).
\btitle{Detection and interpretation of expression quantitative trait loci
  (eQTL)}.
\bjournal{Methods (San Diego, Calif.)}
\bvolume{48}
\bpages{265-76}.
\bdoi{10.1016/j.ymeth.2009.03.004}
\end{barticle}
\endbibitem

\bibitem[\protect\citeauthoryear{Newey and Mcfadden}{1994}]{r46}
\begin{barticle}[author]
\bauthor{\bsnm{Newey},~\bfnm{Whitney}\binits{W.}} \AND
  \bauthor{\bsnm{Mcfadden},~\bfnm{Daniel}\binits{D.}}
(\byear{1994}).
\btitle{Large sample estimation and hypothesis tesing}.
\bjournal{Handbook of Econometrics}
\bvolume{4}
\bpages{2111-2245}.
\bdoi{10.1016/S1573-4412(05)80005-4}
\end{barticle}
\endbibitem

\bibitem[\protect\citeauthoryear{Newey and Powell}{1987}]{r70}
\begin{barticle}[author]
\bauthor{\bsnm{Newey},~\bfnm{Whitney~K}\binits{W.~K.}} \AND
  \bauthor{\bsnm{Powell},~\bfnm{James~L}\binits{J.~L.}}
(\byear{1987}).
\btitle{Asymmetric least squares estimation and testing}.
\bjournal{Econometrica}
\bvolume{55}
\bpages{819-847}.
\end{barticle}
\endbibitem

\bibitem[\protect\citeauthoryear{Pei et~al.}{2018}]{pei2018robust}
\begin{barticle}[author]
\bauthor{\bsnm{Pei},~\bfnm{Huimin}\binits{H.}},
  \bauthor{\bsnm{Wang},~\bfnm{Kuaini}\binits{K.}},
  \bauthor{\bsnm{Lin},~\bfnm{Qiang}\binits{Q.}} \AND
  \bauthor{\bsnm{Zhong},~\bfnm{Ping}\binits{P.}}
(\byear{2018}).
\btitle{Robust semi-supervised extreme learning machine}.
\bjournal{Knowledge-Based Systems}
\bvolume{159}
\bpages{203--220}.
\end{barticle}
\endbibitem

\bibitem[\protect\citeauthoryear{Song, Lin and Zhou}{2023}]{r47}
\begin{barticle}[author]
\bauthor{\bsnm{Song},~\bfnm{Shanshan}\binits{S.}},
  \bauthor{\bsnm{Lin},~\bfnm{Yuanyuan}\binits{Y.}} \AND
  \bauthor{\bsnm{Zhou},~\bfnm{Yong}\binits{Y.}}
(\byear{2023}).
\btitle{A general M-estimation theory in semi-supervised framework}.
\bjournal{Journal of the American Statistical Association}.
\bdoi{10.1080/01621459.2023.2169699}
\end{barticle}
\endbibitem

\bibitem[\protect\citeauthoryear{Sottile et~al.}{2020}]{r51}
\begin{barticle}[author]
\bauthor{\bsnm{Sottile},~\bfnm{Gianluca}\binits{G.}},
  \bauthor{\bsnm{Frumento},~\bfnm{Paolo}\binits{P.}},
  \bauthor{\bsnm{Chiodi},~\bfnm{Marcello}\binits{M.}} \AND
  \bauthor{\bsnm{Bottai},~\bfnm{Matteo}\binits{M.}}
(\byear{2020}).
\btitle{A penalized approach to covariate selection through quantile regression
  coefficient models}.
\bjournal{Statistical Modelling}
\bvolume{20}
\bpages{369-385}.
\bdoi{10.1177/1471082X19825523}
\end{barticle}
\endbibitem

\bibitem[\protect\citeauthoryear{Tauchen}{1985}]{r48}
\begin{barticle}[author]
\bauthor{\bsnm{Tauchen},~\bfnm{George}\binits{G.}}
(\byear{1985}).
\btitle{Diagnostic testing and evaluation of maximum likelihood models}.
\bjournal{Journal of Econometrics}
\bvolume{30}
\bpages{415-443}.
\bdoi{10.1016/0304-4076(85)90149-6}
\end{barticle}
\endbibitem

\bibitem[\protect\citeauthoryear{Ullah, Wang and Yao}{2023}]{r69}
\begin{barticle}[author]
\bauthor{\bsnm{Ullah},~\bfnm{Aman}\binits{A.}},
  \bauthor{\bsnm{Wang},~\bfnm{Tao}\binits{T.}} \AND
  \bauthor{\bsnm{Yao},~\bfnm{Weixin}\binits{W.}}
(\byear{2023}).
\btitle{Semiparametric partially linear varying coefficient modal regression}.
\bjournal{Journal of Econometrics}
\bvolume{235}
\bpages{1001-1026}.
\bdoi{10.1016/j.jeconom.2022.09.002}
\end{barticle}
\endbibitem

\bibitem[\protect\citeauthoryear{Wang and Lian}{2020}]{r68}
\begin{barticle}[author]
\bauthor{\bsnm{Wang},~\bfnm{Lei}\binits{L.}} \AND
  \bauthor{\bsnm{Lian},~\bfnm{Heng}\binits{H.}}
(\byear{2020}).
\btitle{Communication-efficient estimation of high-dimensional quantile
  regression}.
\bjournal{Analysis and Applications}
\bvolume{18}
\bpages{1057-1075}.
\bdoi{10.1142/S0219530520500098}
\end{barticle}
\endbibitem

\bibitem[\protect\citeauthoryear{Wang et~al.}{2019}]{r57}
\begin{barticle}[author]
\bauthor{\bsnm{Wang},~\bfnm{Daixin}\binits{D.}},
  \bauthor{\bsnm{Lin},~\bfnm{Jianbin}\binits{J.}},
  \bauthor{\bsnm{Cui},~\bfnm{Peng}\binits{P.}},
  \bauthor{\bsnm{Jia},~\bfnm{Quanhui}\binits{Q.}},
  \bauthor{\bsnm{Wang},~\bfnm{Zhen}\binits{Z.}},
  \bauthor{\bsnm{Fang},~\bfnm{Yanming}\binits{Y.}},
  \bauthor{\bsnm{Yu},~\bfnm{Quan}\binits{Q.}},
  \bauthor{\bsnm{Zhou},~\bfnm{Jun}\binits{J.}},
  \bauthor{\bsnm{Yang},~\bfnm{Shuang}\binits{S.}} \AND
  \bauthor{\bsnm{Qi},~\bfnm{Yuan}\binits{Y.}}
(\byear{2019}).
\btitle{A semi-supervised graph attentive network for nancial fraud detection}.
\bjournal{2019 IEEE International Conference on Data Mining}
\bpages{598-607}.
\end{barticle}
\endbibitem

\bibitem[\protect\citeauthoryear{Xiong, Zheng and Zhang}{2022}]{r43}
\begin{barticle}[author]
\bauthor{\bsnm{Xiong},~\bfnm{Y}\binits{Y.}},
  \bauthor{\bsnm{Zheng},~\bfnm{Z}\binits{Z.}} \AND
  \bauthor{\bsnm{Zhang},~\bfnm{W}\binits{W.}}
(\byear{2022}).
\btitle{Variable selection in high-dimensional extremile regression via the
  quasi elastic net}.
\bjournal{JUSTC}
\bvolume{52}.
\end{barticle}
\endbibitem

\bibitem[\protect\citeauthoryear{Yuval and Rosset}{2022}]{r55}
\begin{barticle}[author]
\bauthor{\bsnm{Yuval},~\bfnm{O}\binits{O.}} \AND
  \bauthor{\bsnm{Rosset},~\bfnm{S}\binits{S.}}
(\byear{2022}).
\btitle{Semi-supervised empirical risk minimization: Using unlabeled data to
  improve prediction}.
\bjournal{Electronic Journal of Statistics}
\bvolume{16}
\bpages{1434-1460}.
\end{barticle}
\endbibitem

\bibitem[\protect\citeauthoryear{Zhang, Brown and Cai}{2019}]{r58}
\begin{barticle}[author]
\bauthor{\bsnm{Zhang},~\bfnm{Anru}\binits{A.}},
  \bauthor{\bsnm{Brown},~\bfnm{Lawrence}\binits{L.}} \AND
  \bauthor{\bsnm{Cai},~\bfnm{T.}\binits{T.}}
(\byear{2019}).
\btitle{Semi-supervised inference: General theory and estimation of means}.
\bjournal{Annals of Statistics}
\bvolume{47}
\bpages{2538-2566}.
\bdoi{10.1214/18-AOS1756}
\end{barticle}
\endbibitem

\bibitem[\protect\citeauthoryear{Zhang, Cai and Gong}{2023}]{zhang2023semi}
\begin{barticle}[author]
\bauthor{\bsnm{Zhang},~\bfnm{Zijia}\binits{Z.}},
  \bauthor{\bsnm{Cai},~\bfnm{Yaoming}\binits{Y.}} \AND
  \bauthor{\bsnm{Gong},~\bfnm{Wenyin}\binits{W.}}
(\byear{2023}).
\btitle{Semi-supervised learning with graph convolutional extreme learning
  machines}.
\bjournal{Expert Systems with Applications}
\bvolume{213}
\bpages{119164}.
\end{barticle}
\endbibitem

\end{thebibliography}

\end{document}